\documentclass[aps,preprint,pre,showpacs,superscriptaddress,amssymb]{revtex4-2}
\usepackage{amsmath}
\usepackage{caption}
\usepackage{subcaption}
\usepackage{graphicx}  
\usepackage{dcolumn}   
\usepackage{bm}        
\usepackage{relsize}  
\usepackage{verbatim} 
\usepackage{float}
\usepackage{mathtools}
\usepackage{color}
\usepackage{svg}
\usepackage{import}
\usepackage{comment}
\usepackage{soul}

\begin{document}

\title{Lattice Boltzmann Model for Transonic Flows}
 
\author{M. Atif}
\affiliation{Brookhaven National Laboratory, Upton, New York 11973, USA}
\author{N. H. Maruthi}
\author{P. K. Kolluru}
\author{C. Thantanapally }
\affiliation{Sankhyasutra Labs, Bangalore  560064, India}
\author{S. Ansumali}
\affiliation{Jawaharlal Nehru Centre for Advanced Scientific Research, Jakkur, Bangalore  560064, India}

\begin{abstract}
The hydrodynamic limit of a discrete kinetic equation is intrinsically connected with the symmetry of the lattices  used in construction of a discrete velocity model.
On mixed lattices composed of standard lattices the sixth-order (and higher) moment is often not isotropic and thus they are insufficient to  ensure correct imposition of the  hydrodynamic moments.
This makes the task of developing lattice Boltzmann model for transonic flows quite challenging.
We address this by decoupling the physical space lattice from the velocity space lattice to construct a lattice Boltzmann model with very high isotropy. 
The model is entirely on-lattice like the isothermal models, achieves a Mach number of two with only $81$ discrete velocities, and admits a simple generalization  of equilibrium distribution used in isothermal equilibrium.  
We also present a number of realistic benchmark problems to show that the lattice Boltzmann model with a limited number of velocities is not only feasible for transonic flow but is also quite simple and efficient like its subsonic counterpart.
\end{abstract} 

\maketitle

\section{Introduction} 
\label{sec:intro}

The lattice Boltzmann model (LBM), a discrete kinetic scheme, is an efficient method to compute solutions to the Boltzmann equation (BE) in a regime of small deviations from the equilibrium state around rest and isothermal condition described by a fixed reference temperature. 
Even though its standard formulations describe compressible (subsonic and isothermal) fluids on standard lattices (19 or 27 discrete velocities in three-spatial dimensions), the LBM has arguably achieved most of its success and utility in areas with complex and coupled physics in the realm of extreme subsonic and isothermal flows where incompressibility is a good approximation \cite{qian1992lattice,chen1992recovery,chen1998lattice,succi2001lattice,sawant2022consistent,nash2014choice,castro2024lattice,mccullough2024uncertainty}. 
Although the higher-order (multispeed) models have extended the method even for moderate (but still subsonic) Mach number hydrodynamics, LBMs for compressible flows are yet to mature \cite{shan2006kinetic,wilde2020semi,atif2018higher,kolluru2020lattice,coreixas2020compressible,zhao2020toward,feng2019hybrid,saadat2019lattice}. 
Despite recent progress \cite{shan2006kinetic,sofonea2012high,frapolli2014multispeed,atif2018higher}, the same can also be said for LBM with energy conservation when the imposed thermal gradients are strong.   The essential reason for the failure of lower order LBM for strong temperature gradients or high Mach numbers is the perturbative nature of the method and lack of higher order isotropy, which are also its strength at low Mach numbers. 
The simplicity of this procedure and economy in the choice of a number of discrete velocities makes it a method of choice for solving the Boltzmann BGK equation at a low Mach number.  
The challenge is to extend this process for high Mach number flow at a manageable number of discrete velocities with purely Lagrangian on-lattice streaming. 
However, for compressible flow where the Mach number can be order one,  the required number of discrete velocities is often prohibitively large (sometimes as high as $343$), which is a major impediment to the development of LBM as a viable scheme for transonic flows.   

The choice of lattice Boltzmann method for any application crucially depends on two factors: isotropy of the lattice and equilibrium distribution.
Both these essentially originates from the fact that 
in the basic formulation of LBM,  the local equilibrium distribution $f^{\rm eq}\left({\bf v},{\bf u},\theta \right)$ for molecular velocity ${\bf v}$ at fluid velocity $\mathbf{u}$ and temperature $\theta$ is written in  series form in powers of  Mach number (${\rm Ma} = |{\bf u}| / c_s$) where $c_s$ is the sound speed.
This series is expanded around the rest state  Maxwell-Boltzmann distribution $f^{\rm MB}\left({\bf u}={\bf 0}, \theta_0 \right) \equiv  W(\theta_0) $ at a fixed temperature $\theta_0$   (typically  till second order)     as
\begin{align}
\label{eq:MBF}
\begin{split}
f^{\rm eq}\left({\bf v},{\bf u} \right)= W(\theta_0) \left [ 1 +\frac{{\mathbf  u} \cdot {\mathbf  v}}{\theta_0} +\frac{({\mathbf  v }\cdot {\mathbf u})^2 - u^2 \theta_0 }{2\theta_0^2} \right].
\end{split}
\end{align} 
The second order is sufficient to recover the subsonic isothermal limit of hydrodynamics as only the correct form of pressure tensor is needed in this limit \cite{he1998prl}.    
This continuous form of the expanded equilibrium is then evaluated on the quadrature nodes ${\mathbf  c}_i$ to obtain $O({\rm Ma}^3)$ accurate form of discrete equilibrium as 
\begin{equation}
    f^{\rm eq}\left({\mathbf  c}_i, {\bf u} \right)= W_i(\theta_0) \left [ 1 +\frac{{\mathbf  u} \cdot {\mathbf  c}_i}{\theta_0} +\frac{({\mathbf  c }_i\cdot {\mathbf u})^2 - u^2 \theta_0 }{2\theta_0^2} \right],
\end{equation}
where $i$ represents a discrete velocity.
The weights $W_i(\theta_0) \equiv w_i $  ($w_i\geq 0$) for the discrete velocity analogue and second-order accuracy of the pressure tensor get translated as the condition on the isotropy of the velocity set and associated weights as \cite{frisch1986lattice,chen2008discrete}
\begin{equation}
\label{MomentCond4}
   \sum_i w_i=1,\quad   \sum_i w_i  c_{i\alpha} c_{i\beta}=\theta_0\, \delta_{\alpha \beta }, \quad \sum_i w_i    c_{i\alpha} c_{i\beta}  c_{i\gamma} c_{i\kappa}=\theta_0^2\,\Delta_{\alpha \beta \gamma \kappa },
\end{equation}
where $\delta_{\alpha \beta }$ is the Kronecker delta and $\Delta_{\alpha  \beta \gamma \kappa} = \delta_{\alpha\beta}\delta_{\gamma\kappa}+\delta_{\alpha\gamma}\delta_{\beta\kappa}+\delta_{\alpha\kappa}\delta_{\beta\gamma}$ is the isotropic generalized Kronecker delta function.
Imposing these isotropy conditions in itself is a non-trivial task on standard lattices as the fourth-order moment of individual lattices is not isotropic on a lattice of single type \cite{chen2011moment,frisch1986lattice,chen2008discrete,alexander1993lattice}. 
In particular
\begin{align}
\small 
\label{eq:4thisotropy}
\begin{split}
 \sum_{i\in \mathcal{S}}    \frac{c_{i\alpha} c_{i \beta} c_{i \gamma } c_{i \kappa}   }{c^4} &=  2\, \delta_{\alpha  \beta \gamma \kappa  },\\
  \sum_{i\in \mathcal{F}}    \frac{c_{i\alpha} c_{i \beta} c_{i \gamma } c_{i \kappa}   }{c^4}  &= -4 \, \delta_{\alpha  \beta \gamma \kappa  } + 4\, \Delta_{\alpha  \beta \gamma \kappa   },  \\
  \sum_{i\in \mathcal{B}}   \frac{c_{i\alpha} c_{i \beta} c_{i \gamma } c_{i \kappa}   }{c^4}  &  = -16 \, \delta_{\alpha  \beta \gamma \kappa  } + 8 \, \Delta_{\alpha  \beta \gamma \kappa   },
\end{split}
\end{align}
where, $\delta_{\alpha  \beta \gamma \kappa}$ is the anisotropic component leading to a loss of isotropy and $\mathcal{S}, \mathcal{F}, \mathcal{B}$ are sets of discrete velocities in simple cubic (SC), face-centered cubic (FCC), and body-centered cubic (BCC) shells respectively.
Here, we remind that the standard choices of velocity space lattice in three dimensions comprise SC, FCC, and BCC shells [see Appendix \ref{app:shells} for more information on shells and their isotropy conditions derivation]. 
The lattice Boltzmann method in its present form evolved from the observation that in Eq.\eqref{eq:4thisotropy} anisotropic components have an opposite sign, and thus, it is possible to give different weights to various shells to make the resulting combination fully isotropic at the fourth order with  D3Q15, D3Q19 and D3Q27 models being the most prominent ones.  
Indeed, many multispeed lattices also exist where higher isotropy can be obtained with positive weights 
\cite{alexander1993lattice,shan1998discretization,chikatamarla2009lattices,shan2010general,Wahyu2010}.
 


\begin{figure}
    \centering
    \includegraphics[width=0.55\textwidth,trim={0 0cm 0 0cm},clip]{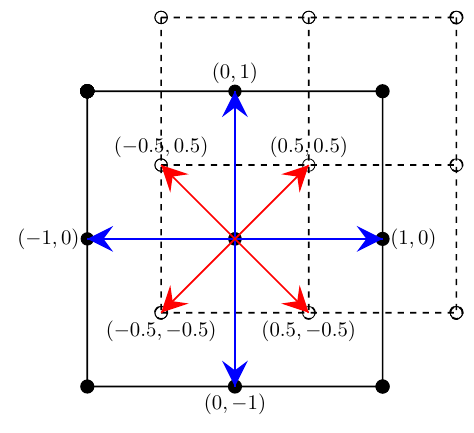}
    \caption{Illustration of a staggered replica grid and half-links that improve the flexibility for imposing required isotropy conditions.}
    \label{fig:replica2d}
\end{figure}

It is evident from the series form in equilibrium in Eq. \eqref{eq:MBF} that one requires higher order contributions for high Mach number case (even at constant temperature). 
However, that often requires matching higher moments of Maxwell--Boltzmann distribution and would thus require more velocity shells \cite{shan2006kinetic,atif2018higher,frapolli2016lattice}. 
Any attempt to model transonic flows  has to ensure that not only the fourth order moment but also the sixth order as well as once contracted $8$th-order moment of the discrete velocity model are isotropic \cite{vahala1998thermal}. 
This is a non-trivial condition as all three fundamental components of the discrete velocity model (SC, FCC, BCC) shells are highly anisotropic at the sixth order. 
This is evident from the explicit expression
\begin{align}
\small
\begin{split}
\label{eq:6thisotropy}
\sum_{i\in \mathcal{S}}    \frac{ c_{i\alpha}   c_{i \beta} c_{i \gamma } c_{i \kappa}  c_{i \eta} c_{i \zeta}}{2\,c^6} &=    \delta_{\alpha  \beta \gamma \kappa  \eta  \zeta}\\
\sum_{i\in \mathcal{F}}    \frac{  c_{i\alpha} c_{i \beta} c_{i \gamma } c_{i \kappa}  c_{i \eta} c_{i \zeta}}{4\,c^6} &=  -13 \, \delta_{\alpha  \beta \gamma \kappa  \eta  \zeta} + \delta_{\alpha  \beta \gamma \kappa  \eta  \zeta}^{(4,2)} \\
\sum_{i\in \mathcal{B}}   \frac{  c_{i\alpha} c_{i \beta} c_{i \gamma } c_{i \kappa}  c_{i \eta} c_{i \zeta}}{8\,c^6} &= 16 \,  \delta_{\alpha  \beta \gamma \kappa  \eta  \zeta} -2 \, \delta_{\alpha  \beta \gamma \kappa  \eta  \zeta}^{(4,2)}  + \Delta_{\alpha  \beta \gamma \kappa  \eta  \zeta},
\end{split}
\end{align}
where $\Delta_{\alpha  \beta \gamma \kappa  \eta  \zeta}$ is the isotropic tensor and $\delta^{(4,2)}_{\alpha  \beta \gamma \kappa  \eta  \zeta}$ is the anisotropic product of Kronecker delta and fourth-order anisotropic tensor.
Similarly, for once contracted 8th moment, we have 
\begin{align}
\small
\begin{split}
\sum_{i\in \mathcal{S}}    \frac{ c_i^2 c_{i\alpha}   c_{i \beta} c_{i \gamma } c_{i \kappa}  c_{i \eta} c_{i \zeta}}{2\,c^8} &=    \delta_{\alpha  \beta \gamma \kappa  \eta  \zeta}\\
\sum_{i\in \mathcal{F}}    \frac{ c_i^2  c_{i\alpha} c_{i \beta} c_{i \gamma } c_{i \kappa}  c_{i \eta} c_{i \zeta}}{8\,c^8} &= -13 \, \delta_{\alpha  \beta \gamma \kappa  \eta  \zeta} + \delta_{\alpha  \beta \gamma \kappa  \eta  \zeta}^{(4,2)}  \\
\sum_{i\in \mathcal{B}}   \frac{c_i^2  c_{i\alpha} c_{i \beta} c_{i \gamma } c_{i \kappa}  c_{i \eta} c_{i \zeta}}{24\,c^8} &=     16 \,  \delta_{\alpha  \beta \gamma \kappa  \eta  \zeta} -2 \, \delta_{\alpha  \beta \gamma \kappa  \eta  \zeta}^{(4,2)}  + \Delta_{\alpha  \beta \gamma \kappa  \eta  \zeta}.
\end{split}
\end{align}
Thus,  we obtain obvious relations to ensure isotropy at various orders listed in Table \ref{tab:isotropy_cond}.
Essential technical difficulty originates from the fact that numerical simplicity demands LBM to work with fixed grid in velocity space which implies that components of ${\bf c}_i$ are integers.  
It is evident from Table \ref{tab:isotropy_cond} that satisfying these complicated constraints is not possible on lower-order lattices and is a challenge on higher-order lattices.
Previous works have demonstrated that introducing a replica grid which allows ${\bf c}_i$ to take values of half  (see Fig. \ref{fig:replica2d}) allows one to construct discrete velocity models that follow isotropy up to a much higher level while retaining the conceptual simplicity of LBM \citep{namburi2016crystallographic,kolluru2020lattice,atif2018higher}.

\begin{table*}[!htp]
    \centering
    \begin{tabular}{c|c}
    $4^{\rm th}$ order isotropy & 
    $2 \sum_{i \in \mathcal{S}} w_{i} c^4_{i}  
    - 4 \sum_{i \in \mathcal{F}} w_{i} c^4_{i} 
    - 16 \sum_{i \in \mathcal{B}} w_{i} c^4_{i} = 0 $  \\
    \hline
    $6^{\rm th}$ order isotropy  & 
    $2 \sum_{i \in \mathcal{S}} w_{i} c^6_{i} 
    - 52 \sum_{i \in \mathcal{F}} w_{i} c^6_{i} 
    + 128 \sum_{i \in \mathcal{B}} w_{i} c^6_{i} = 0 $ \\
    & 
    $4 \sum_{i \in \mathcal{F}} w_{i} c^6_{i}
    - 16 \sum_{i \in \mathcal{B}} w_{i} c^6_{i} = 0 $ \\
    \hline
    Once-contracted $8^{\rm th}$ order isotropy  & 
    $2 \sum_{i \in \mathcal{S}} w_{i} c^8_{i}  
    - 104 \sum_{i \in \mathcal{F}} w_{i} c^8_{i} 
    - 384 \sum_{i \in \mathcal{B}} w_{i} c^8_{i} = 0 $ \\
    & 
    $8 \sum_{i \in \mathcal{F}} w_{i} c^8_{i}
    - 48 \sum_{i \in \mathcal{B}} w_{i} c^8_{i} = 0 $ \\
    \hline
    Thrice-contracted $10^{\rm th}$ order isotropy & 
     $2 \sum_{i \in \mathcal{S}} w_{i} c^{10}_{i}  
    - 32 \sum_{i \in \mathcal{F}} w_{i} c^{10}_{i} 
    - 432 \sum_{i \in \mathcal{B}} w_{i} c^{10}_{i} = 0 $
    \end{tabular}
    \caption{Conditions on the discrete velocities and shells to satisfy isotropy at the various orders. }
    \label{tab:isotropy_cond}
\end{table*}

In this paper, we derive one such higher-order model that satisfies all the constraints listed in Table \ref{tab:isotropy_cond}. 
The key set of ideas used in this work to construct such a higher-order model for transonic flow are:
\begin{itemize}
    \item {\bf Physical space lattice vs velocity space lattice}:  It was first argued in Ref. \cite{cao1997physical} that the lattice in physical space need not be tied up to the lattices created by the link of discrete velocities.  It was pointed out that the constraint on the moment exists only in the velocity space lattice, and the choice of simple cubic lattice in physical space is only needed to preserve the Lagrangian nature of the numerical scheme, which makes it highly efficient and non-dissipative.  
    \citet{namburi2016crystallographic} illustrated that the replica lattice in physical space brings more flexibility and accuracy by allowing $c_i$ to take values of half on a staggered replica grid (see Fig. \ref{fig:replica2d} for illustration).


    
    \item{\bf Entropic equilibrium and temperature-dependent weights}: Energy conserving LBMs require that pressure, stress, and heat fluxes are thermodynamically consistent and remain accurate up to a high order \citep{ansumali2005consistent,atif2018higher}. 
    This requires that the local pressure is the product of local temperature and density and that deviatoric stress and heat fluxes at equilibrium have minimal errors. 
    For large deviations, we  also present an algorithm for efficient evaluation of the true entropic equilibrium which is constructed as the minimizer of the Boltzmann's entropy. 
    We do so by assembling a $5 \times 5$ linear system and solving for the Lagrange multipliers.
    We present a series form of equilibrium which is an intuitive generalization of isothermal equilibrium.  
    A crucial innovation in our approach is the introduction of temperature-dependent weights  $W_i(\theta)$. 
    Existing works use a single expansion in two variables (velocity and temperature) \cite{vahala1998thermal,chikatamarla2006entropy,li2019temperature} whereas our approach first calculates $W_i(\theta)$ as an series expansion in temperature deviation followed by an expansion in velocity. 
    We know from Ref. \citep{atif2018higher} that this is better for flows with strong thermal deviations as it provides accurate initial values of the Lagrange multiplier for solving the linear system.



    \item {\bf Benchmark problems}: We select a broad range of benchmark problems to evaluate the accuracy of the proposed model. We start by testing the theoretical accuracy on well-known setups such as Sod shock tube and Riemann problems and simulate the transonic cavity which is a problem of high practical relevance in the aerodynamic engineering industry.
    
\end{itemize}

The rest of the paper is organized as follows -- in Sec. \ref{sec:lbm} we briefly review the basics of LBM method and the existing higher-order lattices.
In Sec. \ref{sec:model81} we list the constraints, derive the model for transonic flows, and calculate the entropic equilibrium.
This is followed by a rigorous benchmark of the model by simulating challenging transonic flows in Sec. \ref{sec:benchmark}.
Finally, in Sec. \ref{sec:outlook} we provide an outlook by discussing the future of higher-order lattices.

\section{Lattice Boltzmann Method} \label{sec:lbm}

The set of basic variables in the lattice Boltzmann equation consists of $N_d$ discrete populations $f_i({\bf x},t)$ of discrete velocities ${\bf c}_i$ ($i=0,1,2,\cdots N_d-1$) at location $\bf x$ and time $t$.
The hydrodynamic variables, the mass density $\rho$, the fluid velocity ${\bm u}$, and temperature $\theta$, are related to the  the distribution function as 
\begin{equation}
\label{conslaws}
\rho = \left< f, { \tilde 1} \right> \quad \rho u_{\alpha} = \left< f  ,c_{\alpha} \right> , \quad 
\rho u^2 + 3 \rho \theta = \left< f, c^2 \right>,
\end{equation}
where $\bf \tilde 1$ is the unit vector of length $N_d$ and the bilinear action $\left< \phi, \psi \right> = \sum_{i=0}^{N_d-1} \phi_i \psi_i$.
Here, we use a scaled temperature $\theta$
defined in terms of Boltzmann constant $k_B$ and mass of the particle $m$ as $\theta = k_B T /m$ and the discrete velocities ${\bm c}_i$ are in the unit of $\sqrt{k_1 \theta_0}$ where $k_1$ is a lattice-dependent constant and $\theta_0$ is the base temperature of the lattice.  
For these  models, it is important that zero velocity equilibrium at fixed temperature are positive (i.e. the weights $w_i > 0$) and satisfies the condition that their lower order moments are the same as the moments of Maxwell-Boltzmann distribution.
{
As discussed in Sec. \ref{sec:intro}, this can be imposed by demanding isotropy of the lattice and ensuring the traces at the various orders are same as that of Maxwell-Boltzmann distribution.
For example, let us consider the fourth-order isotropy of the the three shells, i.e, $\mathcal{S} = \{(\pm m,0,0), (0,\pm m,0), (0,0,\pm m) \}$, $\mathcal{F} = \{(\pm n,\pm n,0),(\pm n,0,\pm n), (0,\pm n,\pm n) \}$, $\mathcal{B} = \{ (\pm p, \pm p,\pm p) \}$  and weights $w_s$, $w_f$ and $w_b$ respectively. 
Then we have the fourth-order isotropy condition as
\begin{align}
\begin{split}
2\, w_s\, m^4 - 4\,  w_f \, n^4  - 16\,  w_b \, p^4 = 0, 
\label{isotropy4_unit}
\end{split}
\end{align}
and the fourth-order trace from Eq.\eqref{MomentCond4} as
\begin{align}
\begin{split}
\left<w, c^4 \right> = 6 \, w_s\, m^4 + 48 \, w_f  \, n^4 + 72 \, w_b \, p^4 = 15 \theta_0^2.
\label{condLowerLBM}
\end{split}
\end{align}  
In the literature, we frequently find the use of models like $D3Q15$ and $D3Q19$ that have with $m=n=p=1$ and impose only the fourth-order isotropy and trace  (see Table \ref{tab:weightsbasicmodels} for the weights and velocities and Table \ref{tab:mometnssbasicmodels} for the errors in the non-imposed moments). 
It is straightforward to extend the above idea to multispeed lattices that have $m, n, p$ as sets of natural numbers.
It is worth pointing out here that in replica lattices $m, n, p$ can be $0.5, \, 1.5, \, 2.5$ and so on  \citep{namburi2016crystallographic,atif2018higher,kolluru2020lattice}.
Similarly, at the sixth order one obtains the isotropy conditions 
\begin{align}
\begin{split}
2 \, w_s\, m^6 - 52 \,w_{f} \,n^6 + 128\, w_{b}\, p^6 = 0, \\
4\, w_{s}\, m^6 - 16 w_{b} \,n^6 = 0, 
\end{split}    
\end{align}
and the sixth-order trace as
\begin{align}
 \left<w, c^6 \right> = 6 \, w_s\, m^6 + 96 \, w_f  \, n^6 + 216 \, w_b \, p^6 = 105 \theta_0^3.   
 \label{sixthtrace}
\end{align}
It is evident that constructing a discrete velocity set that satisfies Eqs. \eqref{isotropy4_unit} -- \eqref{sixthtrace} is a non-trivial task.
}

The $D3Q27$ model has the full fourth-order as well as one component of the sixth-order moment $\left<w, c_x^2 c_y^2 c_z^2 \right>$ imposed. 
The non-imposed moments at the sixth-order contribute to errors for $D3Q15$ and $D3Q19$ model some of which are absent for the $D3Q27$ model. 
The Zero-One-Three (ZOT) lattice developed by Chikatamarla and Karlin has been frequently used for turbulent flows \cite{chikatamarla2009lattices}. 
The full ZOT lattice has 125 discrete links ($D3Q125$), whereas the pruned one has $41$ discrete links ($D3Q41$). 
The $D3Q41$ lattice imposes the fully-expanded sixth-order tensor, and thus has zero error at the contracted sixth orders, whereas the $D3Q125$ lattice imposes some moments at the eighth and tenth orders too. 
In the aforementioned lattices the moments imposed at the higher order are mostly from the fully expanded set of moments. 
Such lattices work well for turbulent flows but have not yet found widespread acceptance for simulating flows with large temperature variation or at high Mach number. 
It should be noted that all odd moments are zero and that imposing the fully contracted tensor $\left<w, c^{2n}\right>$ for any integer $n$ is equivalent to imposing $\left<w, c^{2(n-1)} c_x^2 \right>$ due to the underlying isotropy and $c^2 = c_x^2 + c_y^2 +c_z^2$. 

To minimize the number of discrete velocities, in the replica lattices the imposed moments are chosen from the contracted set. 
The development of lattices like $RD3Q41$ and $RD3Q67$ emphasized the importance of imposing trace of the higher-order moments.
Here, $R$ represents ``replica" as these lattices make use of staggered grid and contain links with magnitude $0.5$ \cite{namburi2016crystallographic,atif2018higher,kolluru2020lattice}. 
The key idea behind the development of replica lattices was their optimal discretization of position space, with an additional advantage that they offer more freedom to impose moments as these lattices are not derived from the Gauss-Hermite quadrature.
In Ref. \cite{atif2018higher}, it was shown that while the sixth-order isotropy of the reference equilibrium (at zero velocity and reference temperature) is a must, one only requires the trace of the eighth and tenth-order moments to match with their corresponding Maxwell-Boltzmann values for accurately simulating thermal flows. 
Thus, along with the lower order moments, if we demand the isotropy at the sixth order and traces of the sixth and eighth-order moments match the Maxwell-Boltzmann moments
we obtain the $RD3Q41$ lattice which has been found suitable for mildly thermal and weakly-compressible flows.
In addition to the above moments, if we impose isotropy at twice contracted eights order $\sum_i c_i^4 c_{i\alpha} c_{i \beta} c_{i \gamma } c_{i \kappa}$ and traces at the eight order and tenth-order moment we obtain the $RD3Q67$ lattice which remains accurate for temperature deviations as large as $50\%$ from the base temperature $\theta_0$. 

Once the velocity space is discretized into a discrete velocity 
set satisfying the moments of weights as discussed above, one works with the populations $f$ corresponding to each direction as the basic working element. 
The physical space is discretized into a series of grid nodes that are linked together by the discrete velocity directions. 
At any point on the lattice, the neighboring nodes are located at distance of $\Delta x = m \mathbf{c}_i\Delta t$, where $m$ is a natural number \citep{succi2001lattice}. 
This feature of LBM  allows for construction of a computationally attractive algorithm where $f_i$ successively streams along the grid and collides at the nodes. 
Thereafter, the evolution  equation reads 
\begin{equation}
f_i(\mathbf{x}+\mathbf{c}_i\Delta t,t+\Delta t)=f_i(\mathbf{x},t)+ 
\alpha\beta[f_i^{\rm{eq}}(\rho, {\bf u} ) -f_i(\mathbf{x},t)],
\label{discretelbm}
\end{equation} 
where $\alpha=2$ and $\beta={\Delta t }/(2\tau+\Delta t)$ is related to the 
kinematic viscosity $\nu$ via relaxation time $\tau=\nu / \theta_0$, with $\theta_0$ as the reference temperature. 
The above equation can also be derived by trapezoidal integration of discrete velocity model with BGK collision \citep{he1998novel}. 
The entropic formulation of LBM has an extra step where $\alpha$ is found as the root of the entropy estimate
\begin{equation}
H[f(\mathbf{x},t) + \alpha \left( f^{\rm eq}(\rho,{\bm u}, \theta) -f(\mathbf{x},t) \right)] - H[f(\mathbf{x},t)] = 0,
\label{delH}
\end{equation}
where $H$ is a convex entropy function. It restores the thermodynamic consistency embedded in the Boltzmann description \citep{ansumali2005consistent,atif2017,atif2022,hosseini2023entropic}. 
This method ensures $H$ theorem for discrete space-time formulation, and thus leads to a nonlinearly stable solver that is effective in context of flows with sharp gradients. 
Thus, entropic LBM is quite suitable for the case of thermal and compressible flows where gradients can be sharp \citep{mcnamara1997}. 
Various numerical techniques have been proposed
to ensure the correctness and efficient implementation of this step \citep{ansumali2000stabilization,boghosian2001entropic,ansumali2002entropy,tosi2006,chikatamarla2006entropic,ansumali2002single,gorban2012allowed,brownlee2007stability}. 
The closed form analytical expression for $\alpha$ was found in Ref. \citep{atif2017}.


\section{Transonic Model and Discrete Entropic Equilibrium}
\label{sec:model81}

The choice of the discrete equilibrium distribution $f_i^{\rm eq}(\rho, {\bf u}, \theta )$ is considered crucial in LBM. 
It is known that in order to get the correct thermohydrodynamic limit, the moments of equilibrium distribution should match the moments of the Maxwell-Boltzmann distribution \cite{atif2018higher}. The conditions on the moments of the equilibrium translate to constraints on the weights, discrete velocities, and the reference temperature $\theta_0$. 
In general, the higher the order of the moments imposed, the more accurate is the model. 
In the previous section, we listed the constraints on some well-known lattices and two recently developed replica lattices for compressible and thermal flows.
The conditions on isotropy and traces discussed in the previous section are extended to a higher order to design lattices that remain accurate at higher temperature deviations and transonic flows.
The large deviation of stresses in transonic flows require isotropy of the lattice at fourth, sixth, once-contracted eighth order $\sum_i c_i^2 c_{i\alpha} c_{i \beta} c_{i \gamma } c_{i \kappa} c_{i \eta } c_{i \zeta} $, and thrice-contracted tenth-order $\sum_i c_i^6 c_{i\alpha} c_{i \beta} c_{i \gamma } c_{i \kappa}$ explicitly written as
\begin{align}
\begin{split}
2 \sum_{i \in \mathcal{S}} w_{i} c^4_{i}  
    - 4 \sum_{i \in \mathcal{F}} w_{i} c^4_{i} 
    - 16 \sum_{i \in \mathcal{B}} w_{i} c^4_{i} = 0, \\
2 \sum_{i \in \mathcal{S}} w_{i} c^6_{i} 
    - 52 \sum_{i \in \mathcal{F}} w_{i} c^6_{i} 
    + 128 \sum_{i \in \mathcal{B}} w_{i} c^6_{i} = 0, \\
4 \sum_{i \in \mathcal{F}} w_{i} c^6_{i}
    - 16 \sum_{i \in \mathcal{B}} w_{i} c^6_{i} = 0, \\
2 \sum_{i \in \mathcal{S}} w_{i} c^8_{i}  
    - 104 \sum_{i \in \mathcal{F}} w_{i} c^8_{i} 
    - 384 \sum_{i \in \mathcal{B}} w_{i} c^8_{i} = 0, \\
8 \sum_{i \in \mathcal{F}} w_{i} c^8_{i}
    - 48 \sum_{i \in \mathcal{B}} w_{i} c^8_{i} = 0, \\ 
2 \sum_{i \in \mathcal{S}} w_{i} c^{10}_{i}  
    - 32 \sum_{i \in \mathcal{F}} w_{i} c^{10}_{i} 
    - 432 \sum_{i \in \mathcal{B}} w_{i} c^{10}_{i} = 0,    
\end{split}
\end{align}
along with the traces up to the tenth-order moment
\begin{align}
\begin{split}
\left<w,1 \right>=1,  
\left<w, c^2 \right> = 3\, \theta_0,  
\left<w, c^4 \right>= 15 \theta_0^2  \\
\left<w, c^6 \right> = 105 \theta_0^3,  
\left<w, c^8 \right>=  945 \theta_0^4,   
\left<w, c^{10} \right>=  10395 \theta_0^5.
\label{condRed1}
\end{split}
\end{align}  
These conditions can be satisfied on an $81$ link model which is developed in this paper and referred to as $RD3Q81$. 
This model is accurate for transonic as well as strongly thermal flows. 
As the total number of constraints is twelve, we require ten energy shells whose weights along with $w_0$ and $\theta_0$ make a total of twelve unknowns. 
The energy shells chosen are $4$ SC, $2$ FCC and $4$ BCC.
The set of equations has many solutions and we accept the one that satisfies the condition that $w_i>0$ and real.
The discrete velocities and weights corresponding to each shell is listed in the Table \ref{tab:weights}. 


Next, we derive the energy conserving discrete entropic equilibrium as the minimizer of the entropy function.
We begin with the convex entropy function of the Boltzmann form 
\begin{equation}
H= \left< f_i, \log{\left(\frac{f_i}{w_i}\right)}-1 \right>, 
\end{equation}
and construct equilibrium as its minimizer under the constraints of local conservation laws [Eq. \eqref{conslaws}].
The formal solution to the minimization problem is 
\begin{equation}
f_i^{\rm eq}  = w_i \rho  \exp{\left(\mu +\zeta_x c_{ix} + \zeta_y c_{iy} +\zeta_z c_{iz} + \gamma c_{i}^2\right)},
\end{equation}
where $\mu, \zeta_\kappa, \gamma$ are the Lagrange multipliers associated with mass, momentum and energy respectively. 
To obtain the discrete equilibrium in terms of the hydrodynamic fields we need to evaluate the Lagrange multipliers in terms of $\rho, u_\kappa, \theta$. 
These Lagrange multipliers can be evaluated explicitly in terms of conserved moments for the models $D1Q3$, $D2Q9$, $D3Q27$.
For all other models, in practice, they are calculated numerically by first writing the Lagrange multipliers as perturbation series and taking only the leading-order terms, i.e., 
\begin{align}
\begin{split}
f_i^{\rm eq}  = \tilde{f}_i^{\rm eq} \exp  \big( 
 \hat\mu  + \hat\zeta_x c_{ix} + \hat\zeta_y c_{iy} + \hat\zeta_z c_{iz} + \hat\gamma c_{i}^2
\big), 
\label{discentreq}
\end{split}
\end{align}
where $\hat\mu, \hat\zeta_x, \hat\zeta_y, \hat\zeta_z, \hat\gamma$ are the deviations for the Lagrange multipliers and
\begin{equation}
  \tilde{f}_i^{\rm eq} = w_i \rho \exp \left( \mu^{(0)} + \zeta^{(0)}_x c_{ix} + \zeta^{(0)}_y c_{iy} + \zeta^{(0)}_z c_{iz} + \gamma^{(0)} c_{i}^2\right)  
\end{equation}
with the guesses $\mu^{(0)}, \zeta^{(0)}_x, \zeta^{(0)}_y, \zeta^{(0)}_z, \gamma^{(0)}$.
Thereafter, exploiting the approximation $\exp(x) \approx 1 + x$ and imposing the conserved moments one obtains the following system of linear equations
\begin{widetext}
\begin{align}
\begin{bmatrix}
\left<\tilde{f}^{\rm eq}, 1 \right>   & \left<\tilde{f}^{\rm eq}, c_x \right>     & \left<\tilde{f}^{\rm eq}, c_y \right>     & \left<\tilde{f}^{\rm eq}, c_z \right>     & \left<\tilde{f}^{\rm eq}, c^2 \right> \\
\left<\tilde{f}^{\rm eq}, c_x \right> & \left<\tilde{f}^{\rm eq}, c_x^2 \right>   & \left<\tilde{f}^{\rm eq}, c_x c_y \right> & \left<\tilde{f}^{\rm eq}, c_x c_z \right> & \left<\tilde{f}^{\rm eq}, c^2 c_x \right> \\
\left<\tilde{f}^{\rm eq}, c_y \right> & \left<\tilde{f}^{\rm eq}, c_x c_y \right> & \left<\tilde{f}^{\rm eq}, c_y^2  \right>  & \left<\tilde{f}^{\rm eq}, c_y c_z \right> & \left<\tilde{f}^{\rm eq}, c^2 c_y \right> \\
\left<\tilde{f}^{\rm eq}, c_z \right> & \left<\tilde{f}^{\rm eq}, c_x c_z \right> & \left<\tilde{f}^{\rm eq}, c_y c_z \right> & \left<\tilde{f}^{\rm eq}, c_z^2 \right>   & \left<\tilde{f}^{\rm eq}, c^2c_z \right> \\
\left<\tilde{f}^{\rm eq}, c^2 \right> & \left<\tilde{f}^{\rm eq}, c^2 c_x \right> & \left<\tilde{f}^{\rm eq}, c^2 c_y \right> & \left<\tilde{f}^{\rm eq}, c^2 c_z \right> & \left<\tilde{f}^{\rm eq}, c^4 \right> 
\end{bmatrix}
\begin{bmatrix}
\hat\mu \\ \hat\zeta_x \\ \hat\zeta_y \\ \hat\zeta_z \\ \hat\gamma 
\end{bmatrix}
=
\begin{bmatrix}
\rho - \left<\tilde{f}^{\rm eq}, 1 \right> \\ 
\rho u_x - \left<\tilde{f}^{\rm eq}, c_x \right> \\ 
\rho u_y - \left<\tilde{f}^{\rm eq}, c_y \right> \\ 
\rho u_z - \left<\tilde{f}^{\rm eq}, c_z \right> \\ 
3\rho\theta + \rho u^2 - \left<\tilde{f}^{\rm eq}, c^2 \right>
\end{bmatrix}.    
\end{align}
\end{widetext}
Any linear algebra solver can be used for the above system of equations for the Lagrange multipliers $\mu, \zeta_x, \zeta_y, \zeta_z, \gamma$. 
However, the rate of convergence is strongly dependent on the initial guess values of the Lagrange multipliers.
This is where the series form of equilibrium given below becomes useful
\begin{equation}
{\tilde f}_i^{\rm eq }  = {\tilde W_i}(\theta) \left[ 1 + \frac{ u_\alpha {c}_{i\alpha} } {\theta} + \frac{u_\alpha u_\beta}{2 \theta^2} ( {c}_{i\alpha}  {c}_{i\beta}  -  \theta \delta_{\alpha\beta} )     -\frac{5}{3 } \frac{ u^2 }{2 \theta^2} \frac{ \bar{R}}{ 2   +   5 \bar{R}} (c_{i}^2- 3 \theta) + f_i^{(3)} \right],
\end{equation} 
where the series expansion of temperature-dependent weight ${\tilde W_i}(\theta))$ is
\begin{align} 
\begin{split}
\tilde{W}_i(\theta) & =  \rho w_i \biggl[ 1+ \frac{\eta}{2 }\left(   y_i -3 \right)+\frac{\eta^2 } {8}
\left( y_i^2 -10 y_i +15    \right)
\\&+\frac{\eta^3 } {48 }
\left( y_i^3 -21 y_i^2 + 105 y_i-105   \right) + {\cal O}(\eta^4) \biggr].
\end{split}
\end{align}
with the temperature deviation $\eta = \theta/\theta_0 -1$.
Note that the error correction term with ${\bar R}$ will vanish for a model with $12^{\rm th}$-order isotropy.
The full expression with $f_i^{(3)}$ is given in Appendix \ref{app:sereq} where we also compare the errors of the true entropic equilibrium with its  series expansion.
Thus, the series expansion serves as a good initial guess for the Lagrange multipliers and also enables an analysis of the errors associated with the hydrodynamic limit.

\section{Benchmarks}
\label{sec:benchmark}

Several challenging problems in the field of computational gasdynamics have coupled velocity and temperature dynamics with steep gradients.
The complex dynamics has led to the development of a myriad of numerical methods \cite{laneyCGD}.
The aim section of this section is to benchmark the RD3Q81 model for various canonical compressible flows.
The first benchmark problem is the well-known Sod shock tube -- a common test for the accuracy of compressible solvers.
Next, we compare the solutions for two configurations of two-dimensional Riemann problems \cite{lax1998solution}.
We then compare the evolution of kinetic energy and enstrophy for Taylor-Green vortex, followed by an in-depth analysis of transonic flow over an M219 cavity.

\subsection{Sod Shock Tube}

\begin{figure}
    \centering
    \includegraphics[width=0.65\textwidth,trim={0 0cm 0 0cm},clip]{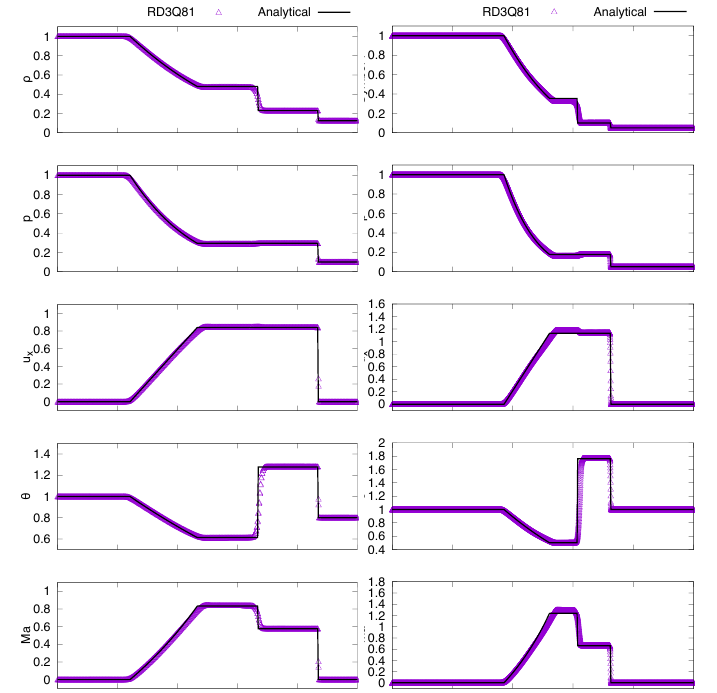}
    \caption{Density, pressure, velocity, temperature, and Ma for Sod shock tube at a Mach number of 0.8 [left] and 1.2 [right].}
    \label{fig:sod-ma0p8}
\end{figure}

The Sod's shock tube problem is the standard test case to check the accuracy and stability of compressible flow solvers.
The existence of analytical solution of the one-dimensional Euler equations makes this problem the first test for compressible flow solvers. 
Here, the domain consists of an initially quiescent fluid divided into two regions $L$ and $R$. 
The two regions, located in $x=-0.5$ to $0.0$ ($L$) and $x=0.0$ to $0.5$ ($R$) respectively, are separated by an interface at $x=0$ across which the initial conditions for density and pressure have a discontinuity.
This sharp discontinuity in at the center of the domain generates a moving compressive shock front in the low density region and rarefaction front in the high density region. 
The two fronts leave behind in the tube a central contact region of uniform 
pressure and velocity \citep{laneyCGD}. 
The left half of the domain is initialized with a density, velocity, and pressure of $\{\rho, u, p\}_L = \{1,0,1\}$ while the right half is varied as
\begin{align}
\begin{split}
{\rm IC}_1: \{\rho, u, p\}_R &= \{0.125,0,0.1\}, \\   
{\rm IC}_2: \{\rho, u, p\}_R &= \{0.05,0,0.05\},   \\ 
{\rm IC}_3: \{\rho, u, p\}_R &= \{0.0125,0,0.0125\},  \\  
{\rm IC}_4: \{\rho, u, p\}_R &= \{0.01,0,0.01\},    
\end{split}    
\end{align}
Here, ${\rm IC}_1$ is the well-known Sod shock setup while the remaining three are its extensions that lead to a higher Mach number in the domain.


\begin{figure}
    \centering
    \includegraphics[width=0.65\textwidth,trim={0 0cm 0 0cm},clip]{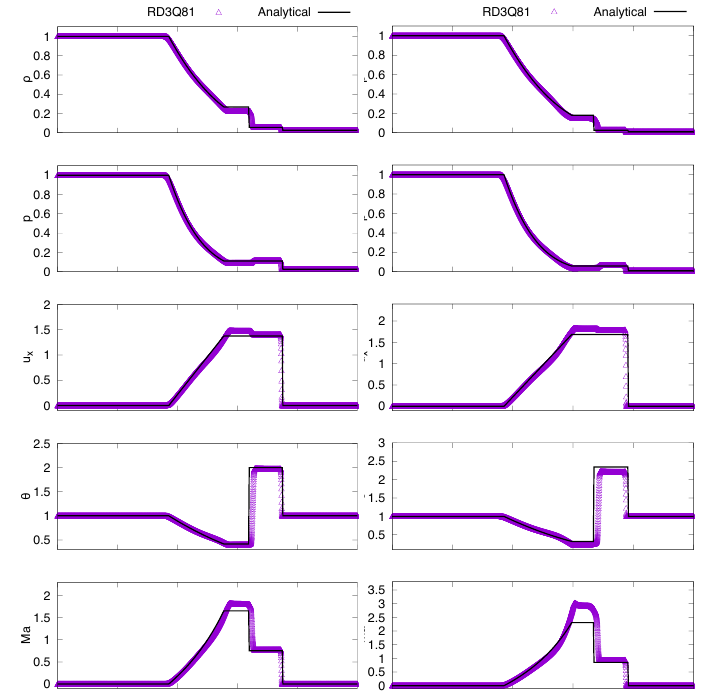}
    \caption{Density, pressure, velocity, temperature, and Ma for Sod shock tube at a Mach number of 1.8 [left] and 2.3 [right].}
    \label{fig:sod-ext}
\end{figure}
 
We discretize the physical domain via a grid of $4000 \times 8 \times 8$ points in the three spatial directions
and use the reference viscosity $\mu=10^{-5}$.
The initial relaxation time $\tau$ is computed from the relation $\mu = \tau p$, where $p$ is $p_R$ or $p_L$ depending on the location of the grid point.
The periodic boundary conditions were implemented in $y,z-$normal directions and standard bounce back in $x$-normal direction. 
The time scale is chosen based on the length of the domain and speed of sound in the right section of the domain. 
The simulations were run till the non-dimensional time $t^*=0.2$, that is  much earlier than either of the fronts hit the wall to avoid any effect of the boundaries.

Figure \ref{fig:sod-ma0p8} contrasts the density, pressure and velocity obtained from the present model with the direct integration of Navier-Stokes-Fourier equations.
It is evident that the speed of the shock is captured accurately by the model for ${\rm IC}_1$ and ${\rm IC}_2$. 
It can be seen that the model remains accurate at a Mach number of 1.3.
The contact region shows a tiny departure from the vertical line which is an artifact of discretization observed in most numerical solvers. 
In Fig. \ref{fig:sod-ext}, we extend the Sod shock problem with initial conditions that give rise of higher Mach numbers (${\rm IC}_3$ and ${\rm IC}_4$) to test the limit of the proposed model.
A small departure from analytical solution is observed at Mach number of 1.8.
At Mach number 2.3, the density and pressure profiles are captured well, however, there is an appreciable deviation velocity and temperature profiles that percolates to a larger error in the predicted Mach number. 
It should however be noted that the model remains numerically stable till a Mach number of 3.

\subsection{Two-dimensional Riemann Problem}

The two-dimensional Riemann problem serves as an important benchmark for numerical methods for compressible flows \citep{lax1998}.
Here, a square domain is decomposed into four quadrants where each quadrant is initialized with a different state.
The resulting dynamics gives rise to complicated flow patterns.
The initial conditions have been compiled into a set of 19 configurations \citep{shulzrinne93} that have been extensively studied by hyperbolic solvers \citep{lax1998,kurganov2000new}.
We select the II and VI configurations to evaluate the accuracy of RD3Q81 model and compare with HyPar \citep{hypar} -- a finite-difference solver for hyperbolic-parabolic partial differential equations.

\begin{figure*}
    \centering
    \includegraphics[width=0.95\textwidth]{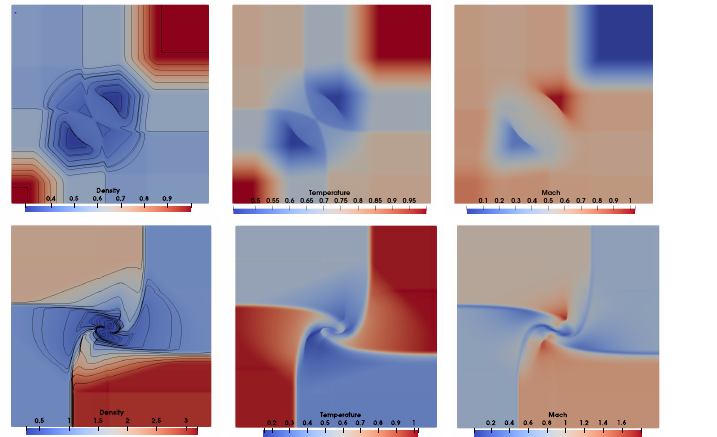}
    \caption{Visualizations of the density, temperature, and Mach for the 2D Riemann setup: configuration II (top row) and VI (bottom row). The black solid lines represent the iso-density contours are included in the first column.}
    \label{fig:case26-riemann2d}
\end{figure*}

\begin{figure}
    \centering
    \includegraphics[width=0.49\textwidth]{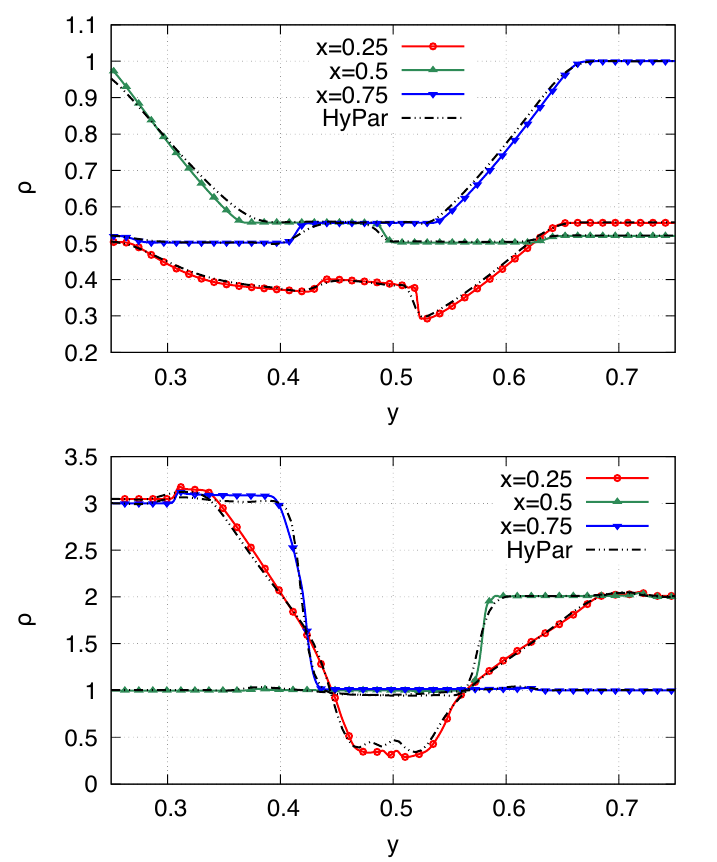}
    \caption{The densities along the three equidistant vertical probes for 2D Riemann configurations II (top) and VI (bottom).}
    \label{fig:riemann-density}
\end{figure}

Figure \ref{fig:case26-riemann2d} visualizes the density, temperature, and Mach numbers for the configuration II (top row) and VI (bottom row).
The presence of shocks and steep-gradients in the domain is captured well by the RD3Q81 model.
The black solid lines represent the iso-density contours are included in the first column.
We perform a quantitative comparison of the density in Fig. \ref{fig:riemann-density}.
The densities along the three equidistant vertical probes  are devoid of any numerical oscillations and exhibit an excellent match with the HyPar's output.

\subsection{Compressible Taylor-Green Vortex}

In this test case we assess the accuracy of the $RD3Q81$ model to capture compressible three-dimensional turbulence accurately. 
The initial condition consists of eight vortices in a triply periodic domain.
The dynamic is well suited to represent the geometry of evolving vortical structures under compressible conditions.
The flow initially transitions to turbulence by creating small vortical structures followed by an isotropic homogeneous decay.
The initial flow field is given by
\begin{align}
u_1 &= V_0 \sin \hat x \, \cos \hat y \cos \hat z, \\	
u_2 &= -V_0 \cos \hat x \sin \hat y \cos \hat z, \\
u_3 &= 0, \\
p &= p_0 + \frac{\rho_0 {V_0}^2}{16} \left[ \cos 2\hat x + \cos 2 \hat y \right] \left[ \cos 2 \hat z + 2 \right], 
\end{align}
where $\hat x = x/L, \hat y = y/L, \hat z = z/L$ and pressure is consistent with the incompressibility condition.
The strength of compressibility of the flow is characterized by the Mach number ${\rm Ma} = V_0 / c_s$ where $c_s = \sqrt{\gamma \theta_0}$ with $\gamma = 5/3$ for an ideal gas.
The Reynolds number ({$\rm Re$}) of the flow is defined by ${\rm Re} = V_0 L/\nu $ where $L = 2\pi$ and $\nu$ is the kinematic viscosity related to the the relaxation time $\tau$ by $\nu = \tau \theta$.

\begin{figure}
    \centering
      \includegraphics[width=0.5\textwidth]{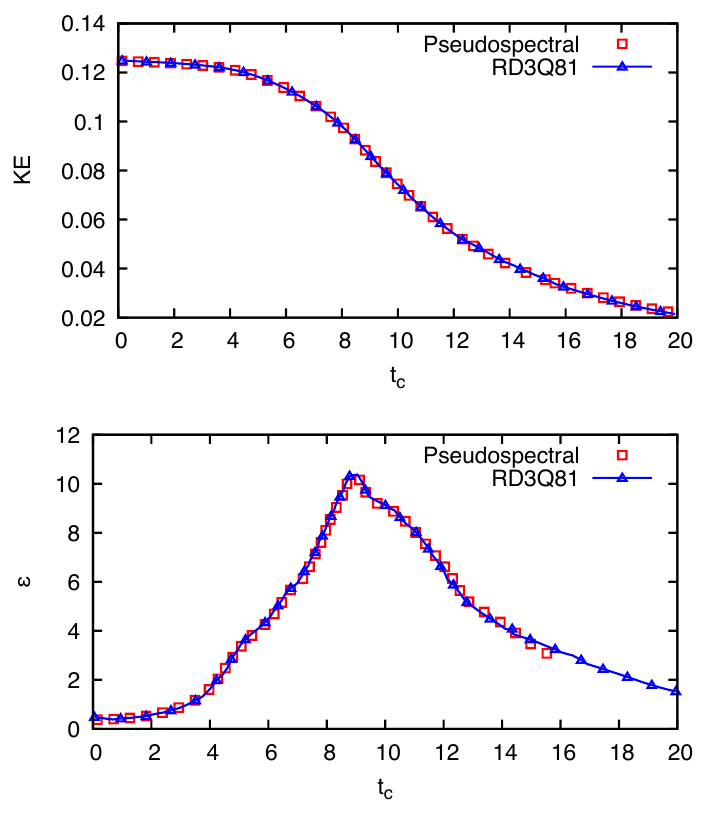}
      \caption{Evolution of kinetic energy (top) and enstrophy (bottom) for ${\rm Ma}=0.1$ and ${\rm Re}=1600$.}
      \label{fig:tgv_ma0p1}
\end{figure}

\begin{figure}
      \includegraphics[width=0.5\textwidth]{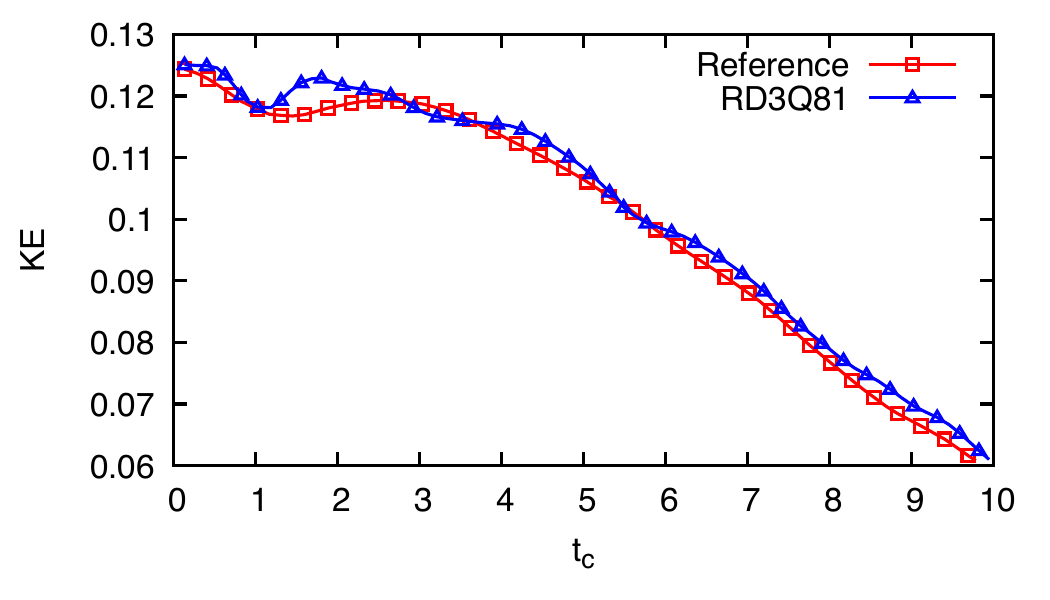}
      \caption{Evolution of kinetic energy for ${\rm Ma}=1$ and ${\rm Re}=400$. 
      }
      \label{fig:tgv_ma1}
\end{figure}

Figure \ref{fig:tgv_ma0p1} plots the kinetic energy and enstrophy for compressible Taylor-Green vortex simulated on a grid of $512^3$ points
at ${\rm Ma}=0.1, {\rm Re}=1600$.
The mean turbulent kinetic energy is computed by
\begin{equation}
{\rm KE} = \frac{1}{ \|\Omega \|} \int_\Omega  \rho \frac {\textbf{u}. \textbf{u}}{2} d\,\Omega,
\end{equation}
where $\rho$ is density, $\Omega$ is volume of the computational domain, and $\textbf{u} = \{u_1, u_2, u_3\}$ is the velocity vector.
The mean enstrophy $\varepsilon$ is given by
\begin{equation}
\varepsilon = \frac{1}{ \|\Omega \|} \int_\Omega  \rho \frac {\omega. \omega}{2} d\,\Omega
\end{equation}
where $\omega = \nabla \times u$ is the vorticity.
We compare the kinetic energy and enstrophy with the output of a pseudospectral solver and observe a good match.
Next, we calculate the kinetic energy at a higher Mach of unity where compressibility effects are more pronounced.
Figure \ref{fig:tgv_ma1} compares the kinetic energy with Ref. \cite{peng2018effects}.


\subsection{Aeroacoustics in a Transonic Cavity}

\begin{figure}
\centering
\includegraphics[width=0.45\textwidth]{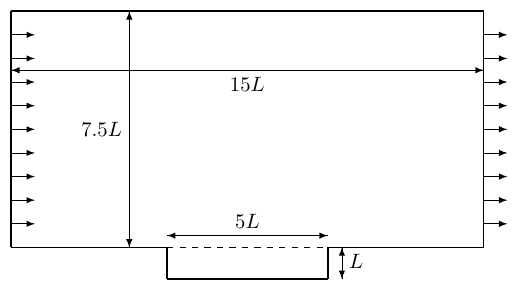}
\caption{Computational setup of the M219 cavity. The spanwise depth is $3L$, with the cavity placed at the center.}
\label{fig:m219_cavity}
\end{figure}

\begin{figure}
\centering
\includegraphics[width=0.45\textwidth]{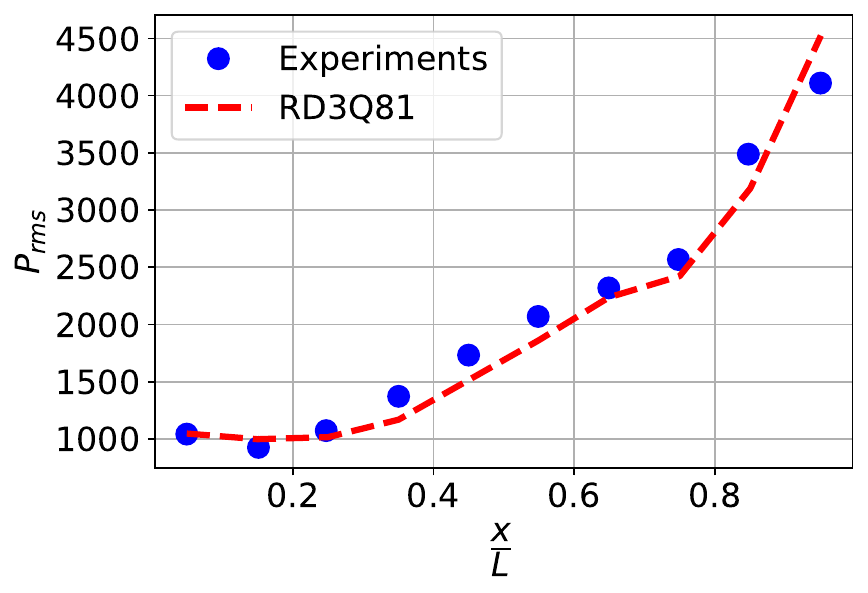}
\caption{Root mean square pressure on the cavity floor.}
\label{fig:prms_floor}
\end{figure}

\begin{figure*}
  \includegraphics[width=0.95\textwidth]{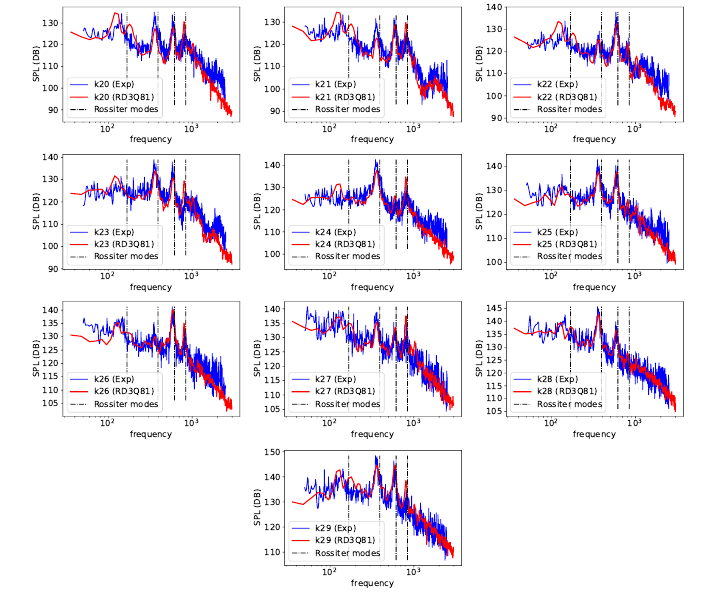}
  \caption{Sound Pressure Level (SPL) at different probe locations ($k20--k29$) on the surface of the cavity}
  \label{fig:spl_1}  
\end{figure*}

As a final benchmark problem of high practical relevance, we analyze the aeroacoustics of M219 cavity flow in this section. 
The flow of open cavity finds applications in various industrial settings such as an open sunroof or a window of a vehicle, landing gear wells or the weapon bays of an aircraft. 
Aeroacoustic investigations are important to reduce undesirable sound and predict structural failures from self-sustaining oscillations in open cavities.
These oscillations arise due to pressure waves generated from the vortical structures generated by the shear layer at the leading edge.
Thus, high-fidelity simulations of cavity acoustics form indispensable tools to understanding of cavity acoustics and designing cavities. 
Such studies offer accurate insights into designs of cavities that lie beyond the  excitation frequencies for various flow configurations. 

\begin{figure*}
  \includegraphics[width=0.75\textwidth]{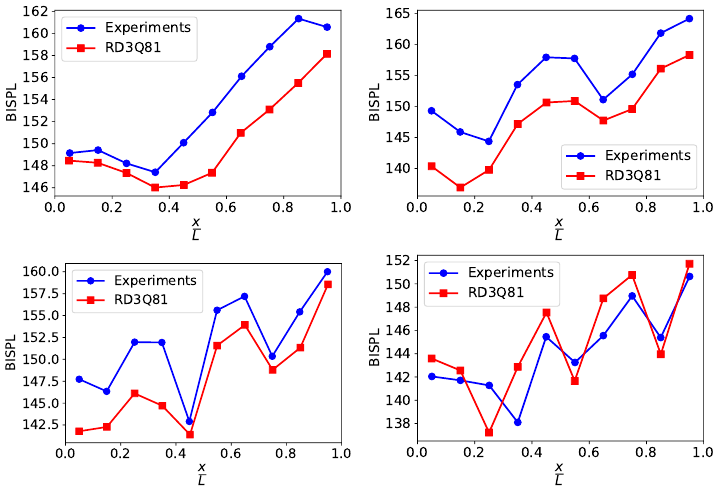}
  \caption{Band Integrated Sound Pressure Level (BISPL) in comparison with experimental results }
\label{fig:bispl_2}
\end{figure*}

Figure \ref{fig:m219_cavity} depicts the computational domain for the cavity in the current investigations.
The cavity has a square cross section of height and depth $L$ which is placed at the center of a domain of spanwise depth $3L$.
The top edge of the domain is placed at a distance of $8.5L$ from the cavity floor.
The length of the cavity is $5L$ placed at the center of a domain of length $15L$.
Thus, both the inlet and the outlet of the computational domain are at a distance of $5L$ from the cavity's leading and trailing edges.
A subsonic inlet boundary condition is specified at the inlet which is based on the total pressure and temperature.
A static pressure boundary condition is applied at the outlet whereas symmetry conditions are applied at the top wall and in spanwise directions.
The diffuse bounce-back boundary condition is specified at the bottom flat plate and the cavity walls \citep{siddharth2014}. 
The numerical simulations are carried out for $L=0.1016 \,{\rm m}$ with a uniform mesh resolution of $0.5 {\rm mm}$ throughout the domain.
The Mach number is 0.85 whereas the Reynolds number based on the cavity length is $6.7 \times 10^6$.
The inlet is maintained at a total pressure of $100969 \,{\rm Pa}$ and temperature of $309 \, {\rm K}$.
In the cavity experiments of Ref. \citep{henshaw2000m219} the pressure was measure with Kulite transducers which were placed at ten equidistant locations (represented by $k20$ to $k29$) between the leading edge and the trailing edge of the cavity.

\begin{table*}
\centering
\caption{Comparison of frequencies of tonal modes between experiments, simulations and Rossiter's formula }
\begin{tabular}{ | l | l | l | l | l |}
\hline
\textbf{Mode frequency (Hz)} & \textbf{Mode 1} & \textbf{Mode 2} & \textbf{Mode 3} &  \textbf{Mode 4} \\ \hline
Experiment & 135 & 350 & 590 & 820  \\	\hline
Modified Rossiter's formula & 167 & 391 & 614 & 838 \\ \hline
RD3Q81 & 133 & 364 & 605 & 810 \\ \hline
$\%$ error between RD3Q81 and experiments        & 1.4   & 4.0  & 2.5  & 1.2 \\ \hline
\end{tabular}
\label{table:modes}
\end{table*}

\begin{table*}
\centering
\caption{Frequency bands for Rossiter modes used for computing BISPL}
\begin{tabular}{ | l | l | l | l | l |}
\hline
\textbf{Rositter mode frequency (Hz)} & \textbf{Mode 1} & \textbf{Mode 2} & \textbf{Mode 3} &  \textbf{Mode 4} \\ \hline
Lower limit & 50 & 250 & 500 & 700  \\	\hline
Upper limit & 250 & 450 & 700 & 900 \\ \hline
\end{tabular}
\label{table:bispl}
\end{table*}

Figure \ref{fig:prms_floor} plots the root mean square of pressure on the cavity floor and compares it with the various probe locations of Ref. \citep{henshaw2000m219}. 
An excellent match can be observed in the figure.
Next, we compute the sound pressure level (SPL) with at the different probe locations and compare them with the experimental data.
A good match is observer in Fig. \ref{fig:spl_1} at all probe locations which also includes the semi-impirical tonal modes proposed by Rossiter \cite{rossiter1964wind}.
Table \ref{table:modes} reports the tonal modes of the experiment and modified-Rossiter's relations \cite{heller1971flow} and compares them with the RD3Q81 model's prediction. It can be seen that the RD3Q81 model is able to accurately predict the tonal modes.

Finally, to understand the mode shapes we compute the band-integrated sound pressure level (BISPL) by integrating the frequencies for each Rossiter mode at all the probe locations.
The lower and upper limits of frequencies chosen for computing BISPL are tabulated in Table \ref{table:bispl} where the width of the frequency band for each mode is 200Hz. 
The mode shapes obtained from BISPL are shown in the Fig. \ref{fig:bispl_2} from where it can be seen from all the BISPL curves closely match the experimental results.

\section{Outlook}
\label{sec:outlook}

In this paper, we have developed a body-centered cubic lattice Boltzmann model $RD3Q81$ suitable for transonic flows.
We have demonstrated that the advantage of these models lies in their conceptual simplicity which is the reason for scalability of lattice Boltzmann method.
We have derived the entropic equilibrium for the model and simulated several benchmark problems of theroretical and practical interest using the proposed model.
This model extends the simplicity of the standard lattice Boltzmann method to flows with moderate Mach numbers where compressibility effects are dominant.
We have also detailed the procedure for constructing such models which relies on imposing higher-order isotropy on the discrete velocity model and matching the trace with Maxwell-Boltzmann distribution up to the tenth order.
The limitation on Mach number is shown to arise due to deviation of the trace of  $12^{\rm th}$ and $14^{\rm th}$ order moments.
However, the procedure can be extended to construct multispeed lattices with 135 and 157 links that impose $12^{\rm th}$ and $14^{\rm th}$ order traces respectively and are suitable for supersonic and hypersonic flows.
It should be noted that the velocities and Mach numbers in the Riemann problems are positive, however, $RD3Q81$ model will attain similar Mach number for the negative velocities also thus extending the range of Mach numbers for realistic applications.
The $RD3Q81$ model is an improvement over $RD3Q67$ model in terms of temperature deviation too, thus, this model will work well for flows with large temperature variations.
For several realistic applications it is important to correctly model the internal degrees of freedom and the specific heat ratio \cite{kolluru2020extended,kolluru2023reduced}, thus in the future we will extend the model to diatomic and polyatomic gases too.

\begin{acknowledgements}
We acknowledge SankhyaSutra Labs and the National Supercomputing Mission facility (Param Yukti) at the Jawaharlal Nehru Center for Advanced Scientific Research for computational resources.  SA  acknowledges support via Abdul Kalam Technology Innovation National Fellowship : 4784 ( INAE/SA/4784 ). 
\end{acknowledgements}

\appendix

\section{Isotropy of the Discrete Velocity Shells in LBM} \label{app:shells}

In this appendix, we derive the isotropy conditions on the shells in LBM.
The three kinds of shells typically used are the simple-cubic ($\mathcal{S}$), face-centered cubic ($\mathcal{F}$), and body-centered cubic ($\mathcal{B}$) shells are sets of discrete velocities as:
\begin{align}
\begin{split}
 \mathcal{S} \, : \, \{ (\pm 1, 0, 0), (0, \pm 1, 0), (0, 0, \pm 1) \},  \\
 \mathcal{F} \, : \, \{ (\pm 1, \pm 1, 0), (0, \pm 1, \pm 1), (\pm 1, 0, \pm 1) \},  \\
 \mathcal{B} \, : \, \{ (\pm 1, \pm 1, \pm 1), (\pm 1, \pm 1, \pm 1)\}.  
\end{split}    
\end{align}
Thus, there are six links in the simple-cubic shell, twelve links in the FCC shell, and eight links in the BCC shell with the directions.
These shells form the basis of all discrete velocity models typically used in LBM.
The multispeed models retain the structure of shells while increasing the magnitudes by an integral factor.

\subsection{Fourth-order isotropy}

We first derive the fourth-order isotropy given in Eq. \eqref{eq:4thisotropy} for $c=1$.
A general fourth-order tensor can be described by a linear combination of isotropic fourth order Kronecker delta tensor [$\Delta_{\alpha\beta\gamma\kappa} = \delta_{\alpha\beta} \delta_{\gamma\kappa} + \delta_{\alpha\gamma} \delta_{\beta\kappa} + \delta_{\beta\gamma} \delta_{\alpha\kappa} $] and anisotropic fourth-order tensor [$\delta_{\alpha\beta\gamma\kappa}$].
Thus, for each shell $\mathcal{S}, \mathcal{F}, \mathcal{B}$ one writes fourth-order tensors representing moments of discrete velocities as 
\begin{align}
\small
\begin{split} \label{fourth1}
\sum_{i\in \mathcal{S}}  \frac{c_{i\alpha} c_{i \beta} c_{i \gamma } c_{i \kappa}   }{c^4} &= a_1 \, \delta_{\alpha\beta\gamma\kappa} + a_2 \, \Delta_{\alpha\beta\gamma\kappa}  ,\\
\sum_{i\in \mathcal{F}}    \frac{c_{i\alpha} c_{i \beta} c_{i \gamma } c_{i \kappa}   }{c^4}  &= a_3 \, \delta_{\alpha\beta\gamma\kappa} + a_4 \, \Delta_{\alpha\beta\gamma\kappa} ,  \\
\sum_{i\in \mathcal{B}}   \frac{c_{i\alpha} c_{i \beta} c_{i \gamma } c_{i \kappa}   }{c^4}  &  = a_5 \, \delta_{\alpha\beta\gamma\kappa} + a_6 \, \Delta_{\alpha\beta\gamma\kappa} .    
\end{split}
\end{align}

Now, we substitute  for $c=1$ and $\alpha = \beta = \gamma = \kappa = x$ in Eq. \eqref{fourth1} to obtain
\begin{align}
\small
\begin{split} \label{fourth2}
\sum_{i\in \mathcal{S}} c^4_{ix} = 2 &= a_1 + 3 a_2, \\
\sum_{i\in \mathcal{F}}  c^4_{ix} = 8  &= a_3  + 3 a_4,  \\
\sum_{i\in \mathcal{B}}  c^4_{ix} = 8  &  = a_5  +  3 a_6.        
\end{split}
\end{align}
and $\alpha = \beta = x,  \gamma = \kappa = y$ in Eq. \eqref{fourth1} to obtain
\begin{align}
\small
\begin{split} \label{fourth3}
\sum_{i\in \mathcal{S}} c^2_{ix} \, c^2_{iy}= 0 &=  a_2, \\
\sum_{i\in \mathcal{F}}  c^2_{ix} \, c^2_{iy} = 4  &=  a_4,  \\
\sum_{i\in \mathcal{B}}  c^2_{ix} \, c^2_{iy} = 8  &  =  a_6.    
\end{split}
\end{align}
The system of equations \eqref{fourth2} -- \eqref{fourth3} are solved to obtain the coefficients of isotropic and anisotropic components of fourth order moments of discrete velocity set.

\subsection{Sixth-order isotropy}

Next, we derive the sixth-order isotropy given in Eq. \eqref{eq:6thisotropy} for $c=1$.
Similar to the procedure in the previous section, we write a generic sixth-order tensor as a linear combination of isotropic tensor $\Delta_{\alpha  \beta \gamma \kappa  \eta  \zeta}$ and anisotropic tensors $\delta_{\alpha  \beta \gamma \kappa  \eta  \zeta}$ and $\delta^{(4,2)}_{\alpha  \beta \gamma \kappa  \eta  \zeta}$. Thus, for each shell one writes sixth-order tensors representing moments of discrete velocities as 

\begin{align}
\small
\begin{split}
\label{sixth1}
\sum_{i\in {\rm SC}}    \frac{ c_{i\alpha}   c_{i \beta} c_{i \gamma } c_{i \kappa}  c_{i \eta} c_{i \zeta}}{c^6} &= a_1 \, \delta_{\alpha  \beta \gamma \kappa  \eta  \zeta} + a_2 \, \delta_{\alpha  \beta \gamma \kappa  \eta  \zeta}^{(4,2)} + a_3 \, \Delta_{\alpha  \beta \gamma \kappa  \eta  \zeta}, \\
\sum_{i\in {\rm FCC}} \frac{ c_{i\alpha}   c_{i \beta} c_{i \gamma } c_{i \kappa}  c_{i \eta} c_{i \zeta}}{c^6}  &= a_4 \, \delta_{\alpha  \beta \gamma \kappa  \eta  \zeta} + a_5 \, \delta_{\alpha  \beta \gamma \kappa  \eta  \zeta}^{(4,2)} + a_6 \, \Delta_{\alpha  \beta \gamma \kappa  \eta  \zeta}, \\
\sum_{i\in {\rm BCC}} \frac{ c_{i\alpha}   c_{i \beta} c_{i \gamma } c_{i \kappa}  c_{i \eta} c_{i \zeta}}{c^6}  &= a_7 \, \delta_{\alpha  \beta \gamma \kappa  \eta  \zeta} + a_8 \, \delta_{\alpha  \beta \gamma \kappa  \eta  \zeta}^{(4,2)} + a_9 \, \Delta_{\alpha  \beta \gamma \kappa  \eta  \zeta}.
\end{split}
\end{align}

Now, we first substitute  $c=1$ and $\alpha = \beta = \gamma = \kappa = \eta = \zeta = x$ in Eq. \eqref{sixth1}
\begin{align}
\small
\begin{split}
\label{sixth2}
\sum_{i\in {\rm SC}}  c_{ix}^6 &= 2 =  a_1 + 15 a_2 + 15a_3\\
\sum_{i\in {\rm FCC}} c_{ix}^6 &= 8 =  a_4 + 15 a_5 + 15 a_6 \\
\sum_{i\in {\rm BCC}} c_{ix}^6 &= 8 =  a_7 + 15 a_8 + 15 a_9,
\end{split}
\end{align}
followed by $\alpha = \beta = \gamma = \kappa = x $ and $= \eta = \zeta = y$ in Eq. \eqref{sixth1}
\begin{align}
\small
\begin{split}
\label{sixth3}
\sum_{i\in {\rm SC}}  c_{ix}^6 &= 0 =  a_2 + 3 a_3\\
\sum_{i\in {\rm FCC}} c_{ix}^6 &= 4 =  a_5 + 3 a_6 \\
\sum_{i\in {\rm BCC}} c_{ix}^6 &= 8 =  a_8 + 3 a_9,
\end{split}
\end{align}
and finally $\alpha = \beta = x $, $\gamma = \kappa = y $, $\eta = \zeta = x$ in Eq. \eqref{sixth1}
\begin{align}
\small
\begin{split}
\label{sixth4}
\sum_{i\in {\rm SC}}  c_{ix}^6 &= 0 = a_3 \\
\sum_{i\in {\rm FCC}} c_{ix}^6 &= 0 = a_5 \\
\sum_{i\in {\rm BCC}} c_{ix}^6 &= 8 = a_9 ,
\end{split}
\end{align}
The system of equations \eqref{sixth2} -- \eqref{sixth4} are solved to obtain the coefficients of isotropic and anisotropic components of sixth order moments of discrete velocity set.

\begin{table*}
\begin{tabular}{|c|c|c|c|c|}
\hline
Discrete velocity & $D3Q15$ & $D3Q19$ & $D3Q27$ & $D3Q41$  \\
\hline
$\theta_0$ & $1/3$ & $1/3$ & $1/3$ & $1-\sqrt{2/5}$ \\
\hline
$(\pm 1,0,0),(0,\pm 1,0),(0,0,\pm 1)$ & $1/9$& $1/18$ & $2/27$ & $37/(5\sqrt{10}) - 91/40$\\
\hline
$(\pm 1,\pm 1,0),(\pm 1,0,\pm 1),(0,\pm 1,\pm 1)$ & - & $1/36$ & $1/54$ & $(55 - 17\sqrt{10})/50$ \\
\hline
$(\pm 1,\pm 1,\pm 1)$ & $1/72$& - & $1/216$ & $(233\sqrt{10} - 730)/1600$\\
\hline
$(\pm 3,0,0),(0,\pm 3,0),(0,0,\pm 3)$ & - & - & - & $(295 - 92\sqrt{10})/16200$ \\
\hline
$(\pm 3,\pm 3,\pm 3)$ & - & - & - & $(130 - 41\sqrt{10})/129600$\\
\hline
\end{tabular}
\caption{Weights corresponding to discrete velocities for the basic LB models at the base temperature $\theta_0$.}
\label{tab:weightsbasicmodels}
\end{table*}

\begin{table*}
\scriptsize
\begin{tabular}{|c|c|c|} \hline
Shells    &  Discrete Velocities($c_i/c$)  &  Weights [$w_i = W(\theta=\theta_0)$] \\ \hline
SC-$1$ & $\left(\pm 1, 0, 0  \right), \left(0, \pm 1, 0  \right), \left( 0, 0, \pm 1 \right) $    & $({388800\,\theta_0 - 2508408\,\theta_0^2 + 7033852\,\theta_0^3 - 7856469\,\theta_0^4 + 2838528\,\theta_0^5 })/{486000}$   \\ \hline
SC-$2$ & $2\left(\pm 1, 0, 0  \right),2 \left(0, \pm 1, 0  \right),2  \left( 0, 0, \pm 1 \right) $    & $({-388800\,\theta_0 + 3091608\,\theta_0^2 - 8269102\,\theta_0^3 + 8694819\,\theta_0^4                       - 2838528\,\theta_0^5})/{3888000}$   \\\hline
SC-$3$ & $3 \left(\pm 1, 0, 0  \right),3 \left(0, \pm 1, 0  \right),3 \left( 0, 0, \pm 1 \right) $  & $({ 388800\,\theta_0 - 3199608\,\theta_0^2 + 8977852\,\theta_0^3 - 9120069\,\theta_0^4 + 2838528\,\theta_0^5})/{30618000}$  \\ \hline
SC-$4$ & $4 \left(\pm 1, 0, 0  \right),4 \left(0, \pm 1, 0  \right),4 \left( 0, 0, \pm 1 \right) $    &$({-97200\,\theta_0 + 809352\,\theta_0^2 - 2315338\,\theta_0^3 + 2396961\,\theta_0^4 - 709632\,\theta_0^5})/{108864000}$    \\ \hline
FCC-$1$ & $ \left(\pm 1, \pm 1, 0  \right),  \left(\pm 1, 0, \pm 1  \right),  \left( 0,\pm 1, \pm 1  \right)$ &  $({8\,\theta_0^3 - 9\,\theta_0^4})/({12})$  \\ \hline
FCC-$2$ & $2  \left(\pm 1, \pm 1, 0  \right),2 \left(\pm 1, 0, \pm 1  \right),  2  \left( 0,\pm 1, \pm 1  \right)$ &   $({-2\,\theta_0^3 + 9\,\theta_0^4})/{768}$ \\ \hline
BCC-${1}/{2}$ & $ \left(\pm 0.5, \pm 0.5,  \pm 0.5  \right)$ & $ ({24300\,\theta_0^2 - 103834\,\theta_0^3 + 126963\,\theta_0^4 - 44352\,\theta_0^5})/{7776}$   \\  \hline
BCC -$1$ & $ \left(\pm 1.0, \pm 1.0,  \pm 1.0  \right)$ &$ ({-24300\,\theta_0^2 + 189022\,\theta_0^3 - 328275\,\theta_0^4 + 177408\,\theta_0^5})/{272160}$   \\  \hline
BCC -$3/2$ & $\left(\pm 1.5, \pm 1.5,  \pm 1.5  \right)$  &
$({2700\,\theta_0^2 - 24778\,\theta_0^3 + 67995\,\theta_0^4 - 44352\,\theta_0^5})/{699840}$   \\  \hline
BCC -$5/2$ & $\left(\pm 2.5, \pm 2.5,  \pm 2.5  \right)$  &  $ ({-972\,\theta_0^2 + 11818\,\theta_0^3 - 43371\,\theta_0^4 + 44352\,\theta_0^5})/{68040000}$ \\  \hline
\end{tabular}
\caption{Energy shells and their corresponding velocities with weights $w_i$ for $RD3Q81$ with $\theta_0 = 0.821035006329447750282591763$.}
\label{tab:weights}
\end{table*}

\begin{table*}
\scriptsize
\begin{tabular}{|c|c|c|c|c|c||c|c|c|}
\hline
Moments & $D3Q15$ & $D3Q19$ & $D3Q27$ & $D3Q41$ & $D3Q125$ & $RD3Q41$ & $RD3Q67$ & $RD3Q81$ \\
\hline
$\theta_0$ & 1/3 & 1/3 & 1/3 & $1-\sqrt{2/5}$ & $1-\sqrt{2/5}$ & 0.29... & 0.74... & 0.82... \\
\hline
${\cal O} (c^4)$ &  &   &  &  &  &  & & \\
\hline
$\left<w, c^2 c_x^2 \right>$ & 0\% & 0\% & 0\%  & 0\% & 0\% & 0\% & 0\% & 0\% \\
\hline
$\left<w, c^2_x c_y^2 \right>$ & Imposed & Imposed & Imposed & Imposed & Imposed & Imposed & Imposed & Imposed  \\
\hline
$\left<w, c^4_x \right>$ & Imposed & Imposed & Imposed & Imposed & Imposed & Imposed & Imposed & Imposed \\
\hline
${\cal O} (c^6)$ &  &  &  &  &  &  &  & \\
\hline
$\left<w, c^4 c_x^2 \right>$ & -5.7\% & -22\% & -17\% & 0\% & 0\% & Imposed & 0\% & 0\%  \\
\hline
$\left<w, c^2 c^4_x \right>$ & -28\% & -28\% & -28\% & 0\% & 0\% & Imposed & 0\% & 0\% \\
\hline
$\left<w, c^2 c^2_x c_y^2 \right>$ & -28\% & -14\% & 0\% & 0\% & 0\% & Imposed & 0\% & 0\% \\
\hline
$\left<w, c_x^2 c_y^2 c_z^2 \right>$ & 200\% & -100\% & Imposed & Imposed & Imposed & 65\% & Imposed & Imposed \\
\hline
$\left<w, c_x^4 c_y^2 \right>$ & 0\% & 0\% & 0\% & Imposed & Imposed &10\% & Imposed & Imposed \\
\hline
$\left<w, c_x^6 \right>$ & 40\% & -40\% & -40\% & Imposed & Imposed & 4.3\% & Imposed & Imposed \\
\hline
${\cal O} (c^8)$ &  &  &  &  &  &  &  & \\
\hline
$\left<w, c^6 c_x^2 \right>$ & -17\% & -51\% & -40\% & 69\% & 32\% & Imposed & Imposed & 0\% \\
\hline
$\left<w, c^4 c^4_x \right>$ & -47\% & -57\% & -53\% & 67\% & 54\% & 4.4\% & Imposed & 0\% \\
\hline
$\left<w, c^4 c^2_x c_y^2 \right>$ & 28\% & -42\% & -19\% & 71\% & 0\% & -6.6\% & Imposed & 0\% \\
\hline
$\left<w, c^2 c_x^2 c_y^2 c_z^2 \right>$ & 200\% & -100\% & 0\% & 218\% & 0\% & 66\% & 40\% & Imposed \\
\hline
$\left<w, c^2 c_x^4 c_y^2 \right>$ & 0\% & -33\% & -22\% & 46\% & 0\% & -18\% & 6.7\% & Imposed\\
\hline
$\left<w, c^2 c_x^6 \right>$ & -66\% & -66\% & -66\% & 76\% & 76\% & 13\% & 2.7\% & Imposed \\
\hline
$\left<w, c_x^4 c^2_y c_z^2 \right>$ & 200\% & -100\% & 0\%  & 218\% & Imposed & 66\% & 40\% & 0\%\\
\hline
$\left<w, c_x^4 c_y^2 c_y^2 \right>$ & 0\% & 0\% & 0\%  & 66\% & Imposed & -6.1\% & 50\% & 33.33\% \\
\hline
$\left<w, c_x^6 c_y^2 \right>$ & -40\% & -40\% & -40\%  & 0\% & Imposed & -43\% & 9.7\% & -19.99\% \\
\hline
$\left<w, c_x^8 \right>$ & -74\% & -74\% & -74\% & 98\% & 98\% & 30\% & 0.70\% & 5.7\% \\
\hline
${\cal O} (c^{10})$ &  &  &  &  &  &  &  & \\
\hline
$\left<w, c^8 c_x^2 \right>$ & -35.32\% & -74.28\% & -61.29\%  & 494\% & 137.1\% & -5.59\% & Imposed &  0\% \\
\hline
$\left<w, c^6 c^4_x \right>$ & -62.33\% & -77.92\% & -72.72\% & 376\% & 208\% & 5.59\%  &  -1.18\% &  Imposed \\
\hline
$\left<w, c^6 c^2_x c_y^2 \right>$ & 5.19\% & -68.83\% & -44.15\% &  670\%  & 29.79\%  & -22.38\% & 1.78\% & Imposed \\
\hline
$\left<w, c^4 c_x^2 c_y^2 c_z^2 \right>$ & 145.4\% & -100\% & -18.18\% & 1658\% & 0\% & 49.64\% & -66.41\% & 17.66\% \\
\hline
$\left<w, c^4 c_x^4 c_y^2 \right>$  & -18.18\% & -63.63\% &  -48.48\% & 506\% & 34.76\% & 34.38\%  & 13.14\% & -2.94\% \\
\hline
$\left<w, c^4 c_x^6 \right>$ & -80\% & -83.63\% & -82.42\% & 324\% & 278\%  & 21.59\% & -6.91\% & 1.17\%  \\
\hline
$\left<w, c^2 c_x^4 c^2_y c_z^2 \right>$ & 145.4\% & -100\% & -18.18\% & 1658\% & 0\% & 49.64\% & 66.41\%  & 17.66\% \\
\hline
$\left<w, c^2 c_x^4 c_y^4 \right>$ & -18.18\% & -45.45\% &  -36.36\% & 516\%  & 0\% & -26.52\% & 64.12\% & 25.97\% \\
\hline
$\left<w, c^2 c_x^6 c_y^2 \right>$ & -50.90\% & -67.27\%  &  61.81\%  & 269\%  & 62.57\%  &  -55.91\% & -1.52\% & -24.41\% \\
\hline
$\left<w, c^2 c_x^8 \right>$ & -88.31\% & -88.31\% &  -88.31\% & 3400\% & 3400\% & 43.73\% & -8.45\% & 8.49\% \\
\hline
$\left<w, c_x^{10} \right>$ & -91.42\% & -91.42\%  & -91.42\% & 393\% &  393\%  & 69.71\% &  -4.91\% & 20.77\% \\
\hline
$\left<w, c_x^8 c_y^2 \right>$ & -74.28\% & -74.28\% &  -74.28\% & 98.32\% & 98.32\% & -73.20\% &  -24.38\% & -46.77\% \\
\hline
$\left<w, c_x^6 c_y^4 \right>$ &  -40\% & -40\% &  -40\%  & 362\% & Imposed & -37.46\% & 76.42\% & 24.18\% \\
\hline
$\left<w, c_x^6 c^2_y c_z^2 \right>$ & 80\% & -100\% & -40\%  & 1189\% & Imposed & 9.73\% & -75.37\% & -13.71\% \\
\hline
$\left<w, c_x^4 c_y^4 c_z^2 \right>$ & 200\% & -100\% & 0\% & 2049\% & Imposed & 82.89\% & -58.95\% & 43.81\% \\
\hline
\end{tabular}
\caption{Percentage errors in the various moments [$\mathcal{O}(c^4)$ to $\mathcal{O}(c^{10})$] of various lattice Boltzmann models. To construct models that remain accurate at higher Mach numbers the errors in the higher moments should be reduced without impacting the lower moments.}
\label{tab:mometnssbasicmodels}
\end{table*}

\section{Series expansion of discrete entropic equilibrium}
\label{app:sereq}

In this section, we first construct the series expansion of temperature-dependent weigths at an arbitrary temperature $\theta$ and then extend it to nonzero velocity equilibrium.

\subsection{Temperature-dependent weights}

\begin{figure}
\includegraphics[scale=0.5]{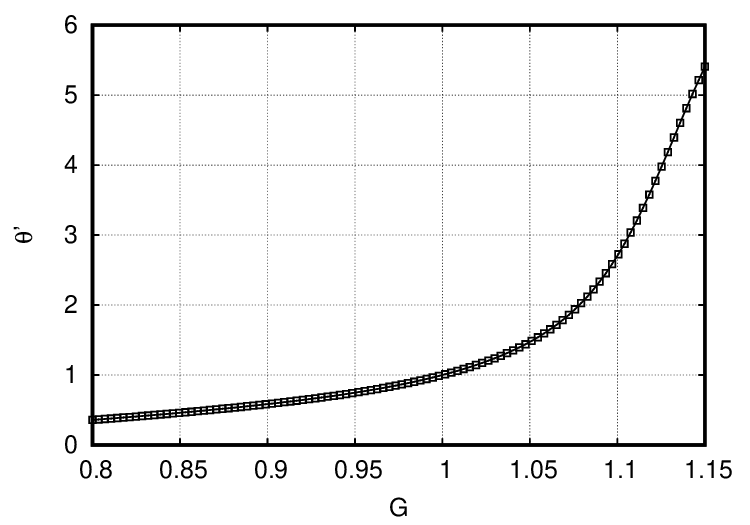}
\caption{The relationship between the temperature 
$\theta$ and its associated Lagrange multiplier.}
\label{fig:theta_G}
\end{figure}

Let the weights of the shells at $\theta = \theta_0$ be
$w_{\rm SC1}$, $w_{\rm SC2}$, $w_{\rm SC3}$, $w_{\rm SC4}$, $w_{\rm FCC1}$, $w_{\rm FCC2}$, $w_{\rm BCCH}$, $w_{\rm BCC1}$, $w_{\rm BCC3H}$, and $w_{\rm BCC5H}$. 
We need to find weights as a functions of temperature as $ W(\theta)$ \cite{ansumali2005consistent}, such that
\begin{equation}
1 = \left< W(\theta), { \tilde 1} \right>, \, 
0 = \left< W(\theta), c_{\alpha} \right> , \, 
3 \theta = \left< W(\theta), c^2 \right>.
\end{equation}
Formally, we can write the temperature-dependent weight as
\begin{equation}
\label{zeroveleq}
W(\theta) = w_i   \exp{\left(\mu^{(0)} + \zeta_\kappa^{(0)} c_\kappa + \gamma^{(0)} c_{i}^2\right)}\equiv  w_i  A G^{4 c_i^2}
\end{equation}
where $\zeta_\kappa^{(0)}= 0$ and $A = \exp{\mu^{(0)}}$ and $G=\exp{\left( \gamma^{(0)} c^2/4\right)}$. Imposing mass and energy conservation one obtains
\begin{align}
\begin{split}
\left< w_i G^{4 c_i^2/c^2}, { \tilde 1} \right> \equiv F_0(G) = \frac{1}{A} \quad {\rm and} \\ \left< w_i G^{4 c_i^2}, c_i^2 \right> \equiv F_2(G) =   \frac{3 \theta^\prime \theta_0}{A},
\end{split}
\end{align}
where $\theta^\prime = \theta/\theta_0$, and $F_0(G), F_2(G)$ are respectively expanded as
\begin{align}
\begin{split}
\label{f0f2geqn}
F_0(G) = w_0 +  6 w_{\rm SC1} G^4 + 6 w_{\rm SC2} G^{16} 
+ 6 w_{\rm SC3} G^{36} 
\\+ 6 w_{\rm SC4} G^{64} + 12 w_{\rm FCC1} G^8 
+ 12 w_{\rm FCC2} G^{32} + 8 w_{\rm BCCH} G^3 
\\+ 8 w_{\rm BCC1} G^{12} + 8 w_{\rm BCC3H} G^{27} 
+ 8 w_{\rm BCC5H} G^{75}, 
\\F_2(G) =  6 w_{\rm SC1} G^4 + 24 w_{\rm SC2} G^{16} + 54 w_{\rm SC3} G^{36} \\+ 96 w_{\rm SC4} G^{64} 
+ 24 w_{\rm FCC1} G^8 + 96 w_{\rm FCC2} G^{32} + 6 w_{\rm BCCH} G^3 
\\+ 24 w_{\rm BCC1} G^{12} + 54  w_{\rm BCC3H} G^{27} 
+ 150 w_{\rm BCC5H} G^{75}.
\end{split}
\end{align}
Simplifying Eq. \eqref{f0f2geqn}, one obtains
\begin{align}
\begin{split}
\theta^\prime(G) = \frac{F_2(G)}{3 \theta_0 F_0(G)}.
\end{split}
\end{align}
The above equation is highly nonlinear in $G$ which represents the relationship between the temperature $\theta$ and its associated Lagrange multiplier (see Fig. \ref{fig:theta_G}). In practice, the temperature $\theta$ at each point is known and $G$ corresponding to it needs to be calculated. This can be accomplished with any iterative numerical solver or by writing a series expansion in the temperature deviation $\eta = \theta/\theta_0 -1$.

To construct the series expansion $\tilde{W}(\theta)$, we write a perturbation series of $W(\theta)$ [Eq. \eqref{zeroveleq}] in $\eta$ around the reference state $\theta=\theta_0$. We expand the Lagrange multipliers $\mu^{(0)}, \gamma^{(0)}$ as
\begin{align}
\begin{split}
 \mu^{(0)} &= \mu^{(0,0)} + \eta \mu^{(0,1)}+  \eta^2 \mu^{(0,2)}+\cdots,\\ 
 \gamma^{(0)} &=\gamma^{(0,0)} + \eta \gamma^{(0,1)}+  \eta^2 \gamma^{(0,2)}+\cdots.
\end{split}
\end{align} 
Substituting the above expansions in Eq. \eqref{zeroveleq} one obtains
\begin{align}
\begin{split}
 \tilde{W}(\theta) =   w_i \exp{\left[ \sum_{k=0}^{\infty} \eta^k \left(\mu^{(0,k)} + \gamma^{(0,k)} c_{i}^2 \right) \right]}.
\end{split}
\end{align} 
Thereafter, using the approximation $\exp{(x)}= 1+x+x^2/2+\cdots$ on the above expression one obtains 
\begin{align}
\begin{split}
 \tilde{W}(\theta)  =  w_i \bigg[ 1 + W^{(0,0)}
 + \eta W^{(0,1)} +\cdots \bigg]
\end{split}
\end{align}
where 
\begin{align}
W^{(0,0)} = \mu^{(0,0)} + \gamma^{(0,0)} c_{i}^2
+ \frac{1}{2}\left(\mu^{(0,0)} + \gamma^{(0,0)} c_{i}^2\right)^2,
\end{align}
and 
\begin{align}
\begin{split}
W^{(0,1)} = \mu^{(0,1)} + \gamma^{(0,1)} c_{i}^2 + \left(\mu^{(0,0)} + \gamma^{(0,0)} c_{i}^2\right) \times \\\left(\mu^{(0,1)} + \gamma^{(0,1)} c_{i}^2\right).
\end{split}
\end{align}

\begin{figure*}
\centering
{\includegraphics[trim={1cm 0 0 0},clip,width=0.75\textwidth]{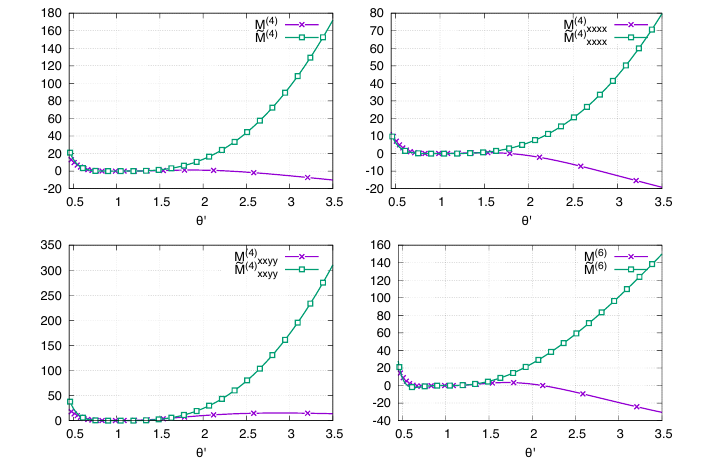}}
\caption{Percentage error in moments of temperature-dependent weights [Eq. \eqref{zeroveleq}] and its series expansion [Eq. \eqref{zeroveleqseries}].}
\label{m_ser_vs_ml}
\end{figure*}   

The conserved moments for $\tilde{W}(\theta)$ are 
\begin{equation}
\left<\tilde{W}(\theta), \left\{1, \frac{c^2}{2} \right\} \right>
= \left\{1, 3\theta_0(1+\eta) \right\}.
\end{equation}
Now, matching the above two moments at various orders of $\eta$ one finds the expressions for the Lagrange multipliers $\mu^{(0,k)}, \gamma^{(0,k)}$.
For example, at ${\cal O}(1)$ one obtains $\mu^{(0,0)} =  \gamma^{(0,0)} = 0$, which is trivial to see as the ${\cal O}(1)$ moments are satisfied by the weight $w_i$. Similarly, at ${\cal O}(\eta)$ one solves
\begin{equation} 
\left<W^{(0,1)}, \left\{1, \frac{c^2}{2} \right\} \right> = \left\{0, 3\theta_0 \eta \right\}.
\end{equation}
to obtain
\begin{equation}
\mu^{(0,1)} = -\frac{3}{2}, \,  \gamma^{(0,1)} = \frac{1}{2 \theta_0}. 
\end{equation}
The final solution reads as
\begin{align}
\begin{split}
\mu^{(0)}=  -\frac{3}{2}\eta +  \frac{6}{8} \eta^2
 -  \frac{1}{2} \eta^3 + \cdots,
\\
\gamma^{(0)} =  \frac{\eta }{2 \theta_0 }  -\frac{\eta ^2}{2 \theta_0 }+\frac{\eta ^3}{2 \theta_0 }   + \cdots.
\end{split}
\end{align} 
Thus, a quartic expression for temperature-dependent weights $\tilde{W}(\theta)$ in terms of $y_i=c_i^2/\theta_0$ is obtained as
\begin{align} 
\begin{split}
\label{zeroveleqseries}
\tilde{W}_i(\theta) & =  \rho w_i \biggl[ 1+ \frac{\eta}{2 }\left(   y_i -3 \right)+\frac{\eta^2 } {8}
\left( y_i^2 -10 y_i +15    \right)
\\&+\frac{\eta^3 } {48 }
\left( y_i^3 -21 y_i^2 + 105 y_i-105   \right) 
\\&+\frac{\eta^4 } {384 }
 \left( y_i^4 -36 y_i^3 + 378 y_i^2-1260 y_i+945  \right)\biggr].
\end{split}
\end{align}
The above expansion is accurate when $\theta$ is close to the base temperature $\theta_0$. The deviation of the moments of the series expansion $\tilde{W}(\theta)$ deviate from the Maxwell-Boltzmann moments at a faster rate than the moments of $W(\theta)$. The practical challenge is to evaluate the Lagrange multiplier $G$ corresponding to the local temperature with maximum accuracy. This can of course be accomplished by an iterative procedure, however, this approach results in an increased computational cost. Therefore, the choice of the initial guess to the iterative solver is important.

In Fig. \ref{m_ser_vs_ml} we plot the errors [( $M / M^{\rm MB} - 1 ) \times 100 $] in fourth-order moments and the trace of the sixth-order moments for the equilibrium given by ${\tilde W}(\theta)$ and $W(\theta)$. The moments are defined as 
\begin{align}
\begin{split}
M^{(4)} = \left< W(\theta), c^2 c^2 \right>, \,     
{\tilde M}^{(4)} = \left< {\tilde W}(\theta), c^2 c^2 \right>, \\
M_{xxxx}^{(4)} = \left< W(\theta), c^2 c^2 \right>, \, 
{\tilde M}_{xxxx}^{(4)} = \left< {\tilde W}(\theta), c_x^2 c_x^2 \right>, \\
M_{xxyy}^{(4)} = \left< W(\theta), c_x^2 c_y^2 \right>, \, 
{\tilde M}_{xxyy}^{(4)} = \left< {\tilde W}(\theta), c_x^2 c_y^2 \right>, \\
M^{(6)} = \left< W(\theta), c^2 c^2 c^2 \right>, \,     
{\tilde M}^{(6)} = \left< {\tilde W}(\theta), c^2 c^2 c^2 \right>, \\
{M}^{\rm (4)MB} = 15 \rho \theta^2, \,
{M}_{xxxx}^{\rm (4)MB} = 3\rho \theta^2, 
{M}_{xxyy}^{\rm (4)MB} = \rho \theta^2, \\
{M}^{\rm (6)MB} = 105 \rho \theta^3.
\end{split}
\end{align}
The above moments are the ones that contribute to the errors at the leading order. It is evident that the moments of the series expansion exhibit faster growth in errors, thus it is important to use $W(\theta)$ evaluated with an iterative numerical procedure for practical applications.

\subsection{Discrete equilibrium at nonzero velocity}

We next derive the full discrete equilibrium distribution at nonzero velocity [Eq. \eqref{discentreq}] by expanding the Lagrange multipliers in $\epsilon$ (representing smallness parameter Mach number) as
\begin{align}
\begin{split}
\mu &=  \mu^{(0)} + { \epsilon} \mu^{(1)}+ {  \epsilon}^2 \mu^{(2)} + \cdots, \\
\zeta_\kappa &= \zeta_\kappa^{(0)} + { \epsilon} \zeta_\kappa^{(1)}+  + {  \epsilon}^2 \zeta_\kappa^{(2)} + \cdots,\\
\gamma &  = \gamma^{(0)} + { \epsilon} \gamma^{(1)}+  + {  \epsilon}^2 \gamma^{(2)} + \cdots,
\end{split}
\end{align}
where $\mu^{(0)}, \zeta_\kappa^{(0)}, \gamma^{(0)}$ are the Lagrange multipliers at zero velocity as discussed in the previous section. Thus, we have 
\begin{align}
\begin{split}
f_i^{\rm eq}  &=  W_i(\theta) \exp{\left[ \sum_{k=1}^{\infty}\epsilon^k Z^{(k)} \right]},
\end{split}
\end{align} 
where $ Z^{(k)}=  \mu^{(k)}+  \zeta_\kappa^{(k)} c_{i\kappa}   + \gamma^{(k)} c_{i}^2 $.
We again use the approximation $\exp{(x)}= 1+x+x^2/2 + \cdots$ on the above expression to obtain a series expansion of the equilibrium as
\begin{equation}
{\tilde f}_i^{\rm eq} = \tilde{W}_i(\theta) \left[ 1 + \epsilon f_i^{(1)} + \epsilon^2 f_i^{(2)} + \epsilon^3 f_i^{(3)} + \cdots \right],    
\end{equation}
where $f_i^{(0)}$ is the zero velocity equilibrium and
\begin{align}
\begin{split}
f_i^{(1)} = Z^{(1)}, \quad
f_i^{(2)} = Z^{(2)} + \frac{1}{2} \left[ Z^{(1)} \right]^2, \\
f_i^{(3)} = Z^{(3)} + Z^{(1)} Z^{(2)} + \frac{1}{6} \left[ Z^{(1)} \right]^3.
\end{split}
\end{align} 
The mass, momentum, and energy conservation constraints for the  discrete equilibrium at any velocity can be written as 
\begin{equation}
\left<{\tilde f}_i^{\rm eq}, \left\{1, c_\alpha, \frac{c^2}{2} \right\} \right>
= \left\{\rho, \epsilon u_\alpha, \epsilon^2 \rho u^2 + 3\rho\theta \right\}.
\end{equation}
The above three conditions are compared at various orders of $\epsilon$ to reveal the discrete equilibrium accurate up to ${\cal O}(\epsilon^3 \eta^4)$ as
\begin{equation}
{\tilde f}_i^{\rm eq }  = {\tilde W_i}(\theta)) \left[ 1 + \frac{ u_\alpha {c}_{i\alpha} } {\theta} + f_i^{(2)} + f_i^{(3)} \right],
\end{equation} 
where
\begin{align}
\begin{split}
f_i^{(2)} &=  \frac{u_\alpha u_\beta}{2 \theta^2} ( {c}_{i\alpha}  {c}_{i\beta}  -  \theta \delta_{\alpha\beta} )     -\frac{5}{3 } \frac{ u^2 }{2 \theta^2} \frac{ \bar{R}}{ 2   +   5 \bar{R}} (c_{i}^2- 3 \theta),
\end{split}
\end{align}
\begin{align}
\begin{split}
f_i^{(3)} &= \frac{u_{\alpha} u_{\beta} u_{\gamma}}{6 \theta^3} \left(   {  c}_{i \alpha} c_{i \beta} c_{i \gamma} -\theta c_{i\kappa}\Delta_{\alpha \beta \gamma \kappa} \right) \\
& \qquad - \left[\frac{5}{6 \theta^3 }  \frac{ \bar{R} u^2 }{ 2   +   5 \bar{R}} \right] u_{\kappa}    { c}_{i \kappa}  (c_{i}^2  - 3 \theta )  \\
& \qquad +\frac{1}{6 \theta^2}  \left(5 u_{\chi} c_{i\chi} u^2  \bar{R}-
   u_{\alpha} u_{\zeta}u_{\eta} c_{i\chi} \bar{R}_{\chi\alpha \zeta \eta} \right),
\end{split}
\end{align}
with
\begin{align}
\bar{R}_{\kappa\alpha \zeta \eta} = \frac{1}{\theta^2}  \left< {\tilde W_i}(\theta)),  c_{i\kappa} c_{i\alpha} c_{i\zeta}c_{i\eta} \right> -  \Delta_{\alpha \kappa \eta \zeta} \\
\qquad {\rm and} \qquad \bar{R} = \frac{1}{15 \theta^2} \left< {\tilde W_i}(\theta)) , c_i^2 c_i^2 \right> -1.
\end{align}
The above equations combined with Eq. \ref{zeroveleqseries} provide a series expansion of the entropic equilibrium at low velocity and temperature deviation. 
It should also be noted that $\bar{R}$ will vanish for a lattice that has the trace at twelfth order imposed.

\bibliographystyle{apsrev4-1}
\bibliography{main}

\end{document}